\documentclass{aa}
\usepackage{txfonts}
\usepackage{graphicx}
\usepackage{mathrsfs}
\usepackage[usenames]{color}
\usepackage{subfigure}

\newcommand{\cL}{\mathcal{L}}
\newcommand{\cM}{\mathcal{M}}
\newcommand{\cH}{\mathcal{H}}

\newcommand{\bfl}{\begin{flushleft}}
\newcommand{\efl}{\end{flushleft}}

\newcommand{\ud}{\mathrm{d}} 
\newcommand{\be}{\begin{equation}}
\newcommand{\ee}{\end{equation}}

\begin{document}

\title{Cosmic-ray ionisation in collapsing clouds}

\author{M. Padovani\inst{1}, P. Hennebelle\inst{2}, and D. Galli\inst{3}}
\authorrunning{M. Padovani et al.}
\titlerunning{Cosmic-ray ionisation in collapsing clouds}

\institute{Laboratoire de Radioastronomie Millim\'etrique, UMR 8112 du CNRS, 
\'Ecole Normale Sup\'erieure et Observatoire de Paris, 24 rue Lhomond, 75231 Paris cedex 05, France\\
\email{padovani@lra.ens.fr}
\and 
CEA, IRFU, SAp, Centre de Saclay, 91191 Gif-Sur-Yvette, France\\
\email{hennebelle@cea.fr}
\and
INAF--Osservatorio Astrofisico di Arcetri, Largo E. Fermi 5, I--50125 
Firenze, Italy\\
\email{galli@arcetri.astro.it}
}


\abstract 
{Cosmic rays play an important role in dense molecular cores, 
affecting their thermal and dynamical evolution and initiating the chemistry.
Several studies have shown that the formation of protostellar discs
in collapsing clouds is severely hampered by the braking torque
exerted by the entrained magnetic field on the infalling gas, as
long as the field remains frozen to the gas.}
{In this paper we examine the possibility that the concentration
and twisting of the field lines in the inner region of collapse can
produce a significant reduction of the ionisation fraction.}
{To check whether the cosmic-ray ionisation rate can fall below the
critical value required to maintain good coupling, we first study
the propagation of cosmic rays in a model of a static magnetised
cloud varying the relative strength of the toroidal/poloidal
components and the mass-to-flux ratio. We then follow the path of
cosmic rays using realistic magnetic field configurations generated
by numerical simulations of a rotating collapsing core with different
initial conditions.}
{We find that an increment of the toroidal component of the magnetic
field, or, in general, a more twisted configuration of the field
lines, results in a decrease in the cosmic-ray flux.  
This is mainly due to the magnetic mirroring effect that is stronger where 
larger variations in the field 
direction are present.
In particular,
we find a decrease of the cosmic-ray ionisation rate below
$10^{-18}$~s$^{-1}$ in the central $300$--$400$~AU, where density
is higher than about $10^{9}$~cm$^{-3}$. This very low value of the
ionisation rate is attained in the cases of intermediate and low
magnetisation (mass-to-flux ratio $\lambda=5$ and 17, respectively)
and for toroidal fields larger than about 40\% of the total field.}
{Magnetic field effects can significantly reduce the ionisation
fraction in collapsing clouds.  We provide a handy fitting formula
to compute approximately the attenuation of the cosmic-ray ionisation
rate in a molecular cloud as a function of the density and the magnetic
configuration.}


\maketitle

\section{Introduction}
\label{intro}

The study of the interaction of cosmic rays (CRs) with the interstellar
matter is a multi-disciplinary task that involves the analysis of
several physical and chemical processes: ionisation of atomic and
molecular hydrogen, energy loss by elastic and inelastic collisions,
energy deposition by primary and secondary electrons, $\gamma$-ray
production by pion decay, generation of small-scale turbulence by
streaming instabilities, and the production of light elements by
spallation reactions. CR ionisation activates the rich chemistry
of dense molecular clouds and determines the degree of coupling of
the gas with the local magnetic field, which in turn controls the
ambipolar diffusion timescale and the star-formation efficiency of a molecular
cloud.

In recent years a wealth of observations from the ground and from
space has provided information and constraints on the flux and the
ionisation rate of cosmic rays.  Detections of large abundances of
H$_{3}^{+}$ in diffuse clouds (e.g. Indriolo et al.~\cite{im12}),
observations of OH$^{+}$ and H$_{2}$O$^{+}$ in low H$_{2}$ fraction
regions (Neufeld et al.~\cite{n10}, Gerin et al.~\cite{g10}),
estimates of enhanced values of the CR ionisation rate in a molecular
cloud close to a supernova remnant (Ceccarelli et al.~\cite{ch11})
as well as the measurement of the $\gamma$ luminosity of molecular      
clouds (e.g. Montmerle~\cite{m10}) raised the questions about the
origin of the CR flux that generates such a high ionisation rate
and how to reconcile these values with those ones measured in denser
clouds that are more than one order of magnitude lower. A number
of studies approached this problem using different strategies,
analysing the effects of Alfv\'en waves on CR streaming (Skilling
\& Strong~\cite{ss76}, Hartquist et al.~\cite{hd78}, Padoan \&
Scalo~\cite{ps05}, Rimmer et al.~\cite{rh12}), magnetic mirroring
and focusing (Cesarsky \& V\"olk~\cite{cv78}, Chandran~\cite{c00},
Padovani \& Galli~\cite{pg11}), or the possible existence of a
low-energy flux of CR particles able to ionise diffuse but not dense
clouds (Takayanagi~\cite{t73}, Umebayashi \& Nakano~\cite{un81},
McCall et al.~\cite{mh03}, Padovani et al.~\cite{pgg09}).

Disc formation is another integral aspect of star formation. One
of the main concerns is the so-called ``catastrophic magnetic braking
problem'' that suppresses the formation of a rotationally supported
disc in the ideal MHD limit during the protostellar accretion phase of a
low-mass forming star (Allen et al.~\cite{al03}, Galli et
al.~\cite{gl06}, Mellon \& Li~\cite{ml08}, Hennebelle \&
Fromang~\cite{hf08}).  Given the observational evidence of discs
on 100~AU or even larger scales at least around Class~I--II protostars,
a number of possible solutions to the problem of catastrophic
magnetic braking have been invoked, including: ({\it i}\/)
non-ideal MHD effects such as Ohmic dissipation (Shu et al.~\cite{sg06},
Dapp \& Basu~\cite{db10}) and Hall effect (Krasnopolsky et
al.~\cite{kl11}, Braiding \& Wardle~\cite{bw12a,bw12b}); ({\it ii})
the possible misalignment between the rotation axis and the magnetic
field direction that acts reducing the braking torque (Hennebelle
\& Ciardi~\cite{hc09}); ({\it iii}\/) depletion of the infalling
envelope that anchors the magnetic field braking (Mellon \&
Li~\cite{ml09}, Machida et al.~\cite{mi11}); ({\it iv}\/) turbulent
diffusion of the magnetic field (Seifried et al.~\cite{sb12},
Santos-Lima et al.~\cite{sd13}). 

Mellon \& Li~(2009) advanced the interesting possibility that a
reduction of the CR ionisation rate, $\zeta^{{\rm H}_2}$, corresponding to a decrease  
of the ionisation fraction by a factor $\sim \sqrt{\zeta^{{\rm H}_2}}$, could result
in sufficient ambipolar diffusion to allow the formation of a rotationally supported disc.
They concluded that both a suppression of the CR flux and a
low level of magnetisation (measured by the non-dimensional mass-to-flux
ratio $\lambda$) were needed in order to circumvent the
magnetic braking problem. Although they did not perform a detailed
exploration of the parameter space of their models, Mellon \& Li~(2009)
found that a value of $\zeta^{{\rm H}_2}=10^{-18}$~s$^{-1}$ was needed to 
spin up the gas significantly during collapse if $\lambda=4$ (but no disc larger than
$\sim 10$~AU was formed in this case), or to form a fully rotationally supported
disc of radius $\sim 50$~AU if $\lambda=13.3$.

Following the suggestion by Mellon \& Li~(2009), in this paper 
we focus on the influence of different magnetic field
and density configurations on the CR propagation following the
conclusions achieved in Padovani et al.~(\cite{pgg09},~\cite{pg11}).
Our aim is to show that
in the inner regions of a cloud, where the disc is formed, magnetic
and column-density effects can indeed cause a significant decrease of the 
interstellar CR ionisation
rate and consequently of the ionisation degree, 
helping to decouple the gas from the magnetic field.

The paper is organised as follows. In Section~\ref{method} we
provide a detailed description of the method used to calculate the
CR ionisation rate. In Section~\ref{semi-analytical} we analyse a
semi-analytical model of a singular isothermal toroid threaded by
a toroidal magnetic field with the purpose of understanding the
role of column density versus magnetic effects. In Section~\ref{nummod}
we explore the evolution of the CR ionisation rate on the initial
conditions (mass-to-flux ratio and alignment between rotation axis
and magnetic field direction) for a number of numerical simulations.
In Section~\ref{comparisonUN} we discuss the variations of the CR
ionisation rate in discs and in Section~\ref{fittingformula} we
give a fitting formula to compute the CR ionisation rate accounting
for the magnetic field configuration.  In Section~\ref{conclusions}
we summarise our conclusions.  Comments on other possible models
of the CR ionisation rate are provided in Appendix~A.

\section{Method}
\label{method}

Cosmic rays, being charged particles, perform an helicoidal motion
around the magnetic field lines and we follow their path starting
from the outer boundary of the computational domain throughout the
core.  
As it is well known, a charged particle of mass $m$ and velocity $v$ spiraling along a magnetic 
field of increasing strength $B$ must increase its pitch angle $\alpha$
(the angle between the particle's velocity and the field direction)
as a consequence of the conservation of kinetic energy
$E_{\rm kin}=(\gamma-1) mv^{2}$ and magnetic moment $\mu=\gamma
mv^{2}\sin^{2}\alpha/2|{\bf B}|$. In particular, for a particle starting from
the intercloud medium (ICM) with a pitch angle $\alpha_{\rm ICM}$ and 
a magnetic field strength $B_{\rm ICM}$, the pitch angle is given by
\be
\label{alpha}
\alpha = \arccos\sqrt{1-\chi+\chi\cos^{2}\alpha_{\rm ICM}}\,,
\ee
where $\chi=|\mathbf{B}|/B_{\rm ICM}$ is the focusing factor
(see e.g. Desch et al.~\cite{dc04} and Padovani \& Galli~\cite{pg11}, hereafter PG11).
The assumption of energy conservation along the particle's trajectory is clearly violated in the 
presence of collisional losses. In principle, Eq.~\ref{alpha} should be replaced by an equation for 
the time evolution of the pitch angle $\alpha$, including the effect of the magnetic field as well 
as the diffusion induced by collisional ionisation of H$_2$ molecules. In the present study we neglect 
these aspects, and while we assume that kinetic energy is conserved along each individual trajectory,
we take into account energy losses globally by propagating the CR spectrum inside the cloud. We have 
verified that this approximation is valid for pitch angles not too small (that evolve to $90^{\circ}$ 
before substantial energy losses occur) and for proton energies larger than about 10~MeV. 
For CR protons of lower energies, 
our treatment overestimates somewhat the efficiency of magnetic mirroring. Since the 
bulk of ionisation from CR protons at the typical column densities of molecular clouds is produced by 
particles in the energy range 1-300~MeV (Padovani et al.~\cite{pgg09}, hereafter PGG09), 
our approximation is satisfactory. For 
CR electrons our approximation is satisfied for all energies of interest.
We also assume that
the number of particles is conserved, ignoring electron capture reactions of CR protons with 
H$_{2}$ and He as well as the $\alpha+\alpha$ fusion reactions that form $^{6}$Li and $^{7}$Li,
because of the small cross sections (Meneguzzi et al.~\cite{ma71}).

The column density of H$_2$ passed through by the particle is given by
\be
N(\alpha)=\int_{0}^{\ell_{\rm max}(\alpha)}n(\ell)\,\ud\ell\,,
\ee
where $\ell_{\rm max}$ is the maximum depth reached inside the core
and $n(\ell)$ is the H$_{2}$ volume density. If the pitch angle
$\alpha$ remains smaller than $\pi/2$ along the entire particle's
trajectory, CRs of sufficient energy can cross the whole core. Vice
versa, if $\alpha$ reaches $\pi/2$ at some point inside the cloud,
the particle is mirrored and it will follow the same field line
backwards.

\begin{figure}[!h]
\begin{center}
\includegraphics[width=.5\textwidth, clip=true]{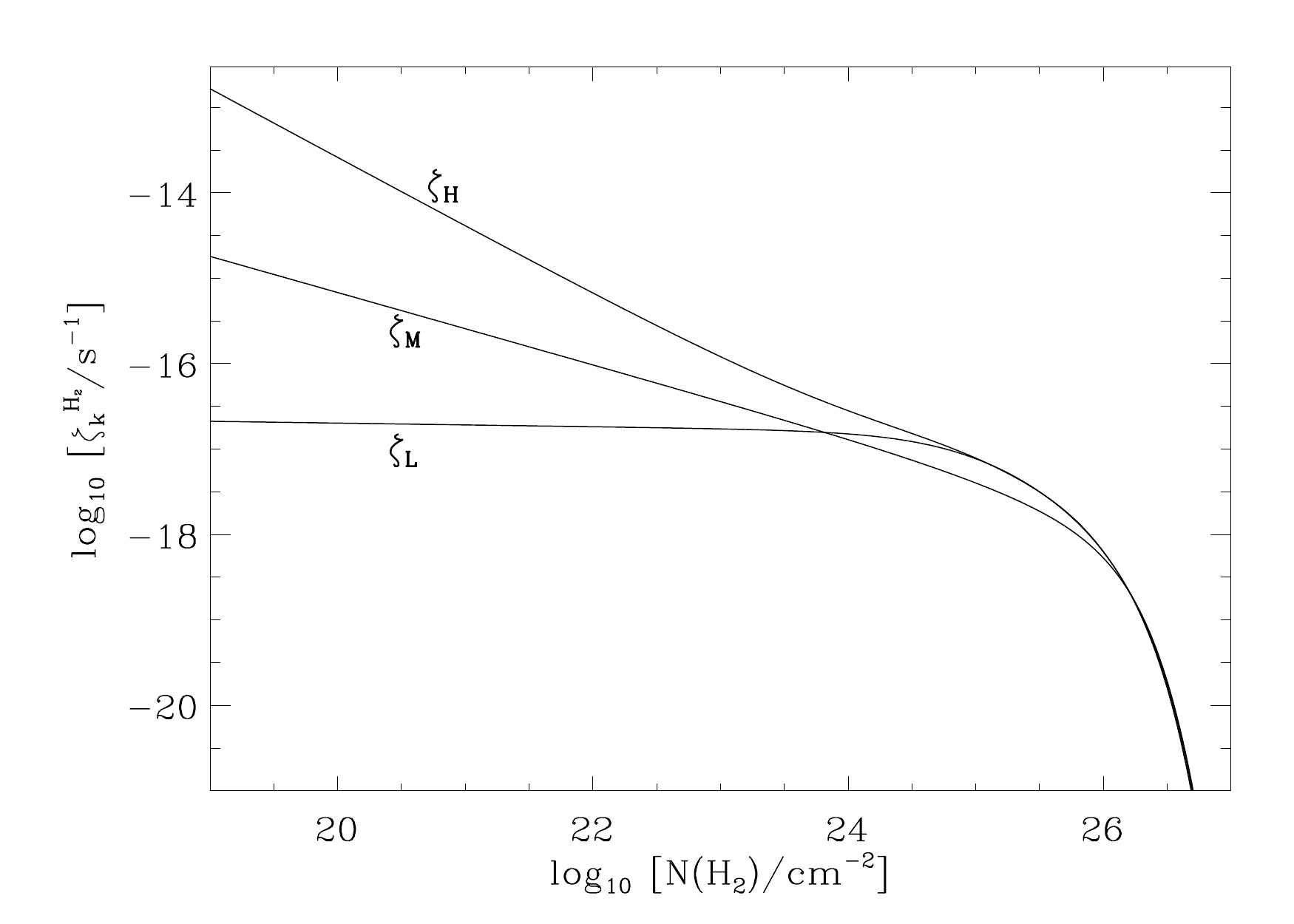}
\caption[]{CR ionisation rate as a function of the molecular hydrogen column
density for the three models described in the text.}
\label{modelfitsAMRpaper}
\end{center}
\end{figure}

Once the column density is known for each value of the initial pitch
angle $\alpha_{\rm ICM}\in[0,\pi/2)$ and for each field line, we
compute the CR ionisation rate $\zeta^{{\rm H}_2}$ using the CR
propagation theory developed in PGG09. 
Since two of their models, those where the ionisation
degree is dominated by CR electrons, are very similar, we assume
three possible trends for the ionisation rate as a function of the
column density, indicated by $\zeta_k$ with $k={\cL}$, ${\cM}$, and ${\cH}$, 
where $\cL$={\em low}, $\cM$={\em medium}, and $\cH$={\em high} 
correspond to the models W98+C00, M02+C00, and the average between
the two models W98+E00 and M02+E00 detailed in PGG09, respectively. 
PGG09 shows that the decrease of the CR ionisation rate with
increasing penetration in the cloud can be described by a power-law
at column densities $N({\rm H}_2)$ in the range $ \sim10^{20}-10^{25}$~cm$^{-2}$,
\be
\zeta^{\rm(low~N)}_{k}\approx \zeta_{0,k}^{(\rm low~N)}
\left[\frac{N(\mathrm{H_{2}})}{10^{20}~\mathrm{cm^{-2}}}\right]^{-a}\,,
\ee
and by an exponential attenuation for $N(\mathrm{H_{2}})\gtrsim10^{25}$~cm$^{-2}$,
\be
\zeta^{\rm(high~N)}_{k}\approx \zeta_{0,k}^{(\rm high~N)}\exp
\left[-\frac{\Sigma(\mathrm{H_{2}})}{\Sigma_{0,k}}\right]\,.
\ee
Padovani et al.~(\cite{pgg13}, hereafter PGG13) give a simple fitting formula
which combines in a single expression the low- and high-column density 
approximations above, 
\be
\label{fitfor}
\zeta_{k}^{\rm H_{2}}(\alpha)=\frac{\zeta_{0,k}^{(\rm low~N)}\zeta_{0,k}^{(\rm high~N)}}%
{\zeta_{0,k}^{(\rm high~N)}\left[\displaystyle{\frac{N(\alpha)}{10^{20}~{\rm cm}^{-2}}}\right]^{a}+%
\zeta_{0,k}^{(\rm low~N)}\left[\exp\left(\displaystyle{\frac{\Sigma(\alpha)}{\Sigma_{0,k}}}\right)-1\right]}\,,
\ee 
where $\Sigma(\alpha)=\mu m_{\mathrm p}N(\alpha)/\cos\alpha$
is the effective surface density seen by a CR propagating with pitch
angle $\alpha$, $m_{\mathrm p}$ the proton mass and $\mu = 2.36$
the molecular weight for the assumed fractional abundances of H$_{2}$
and He. We have used this fitting formula for the three models adopted 
in this work (see Fig.~\ref{modelfitsAMRpaper}).
The fitting coefficients of Eq.~(\ref{fitfor}) are given
in PGG09 and PGG13.

In order to evaluate the ionisation rate along a field line, we
average the contribution of all the particles with different initial
pitch angles over the solid angle
\begin{eqnarray}
\zeta_k^{\rm H_{2}} &=& \frac{\int\zeta_k^{\rm H_{2}}(\alpha)\,\ud\Omega_{\alpha}}{\int\ud\Omega_{\alpha}}=
\frac{2\pi\int_{0}^{\pi/2}\zeta_k^{\rm H_{2}}(\alpha)\sin\alpha\,\ud\alpha}%
{2\pi\int_{0}^{\pi/2}\sin\alpha\,\ud\alpha}=\nonumber\\
&=&\int_{0}^{\pi/2}\zeta_k^{\rm H_{2}}(\alpha)\sin\alpha\,\ud\alpha
\end{eqnarray}
Finally, to account for magnetic focusing, we multiply $\zeta_{k}^{\rm H_{2}}$
by $\chi$. In the following, we choose as a ``fiducial'' spectrum the one
corresponding to the case $k=\cM$ and we neglect the subscript
$\cM$. Appendix~\ref{caseLandH} shows the results for the
other two models $k=\cL,\cH$.

\section{Semi-analytic model}
\label{semi-analytical}

In PG11, we modelled molecular cloud cores
as singular isothermal toroids, namely scale-free, axisymmetric
equilibrium configurations of an isothermal gas cloud under the influence
of self-gravity, gas pressure, and magnetic forces (Li \& Shu~\cite{ls96})
These models are uniquely characterised by the mass-to-flux ratio $\lambda$ 
defined by
\be
\lambda = 2\pi G^{1/2}\frac{M(\Phi)}{\Phi}\,,
\ee
where $G$ is the gravitational constant, $\Phi$ the magnetic flux, and $M(\Phi)$ 
the mass contained in the flux tube $\Phi$.
In this model the purely poloidal magnetic field threading the core takes the
form
\be
\label{Bfield}
\mathbf{B} = \frac{1}{2\pi}\nabla\times
\left(\frac{\Phi(r,\theta)}{r\sin\theta}\mathbf{\hat e}_{\varphi}\right)\,,
\ee
where $\Phi(r,\theta)$ is the magnetic flux function. Separation of variables
requires the equilibrium solution to take the self-similar form
\be\label{FI}
\Phi(r,\theta)=\frac{4\pi c_{s}^{2}r}{G^{1/2}}\phi(\theta)\,,
\ee
where $c_{s}$ is the sound speed and $\phi(\theta)$ is a dimensionless
function.

If the core rotates with respect to the ambient medium, a toroidal component of the magnetic
field is expected to arise. 
We introduce a time-independent toroidal magnetic field component by
adding a term
\be\label{Btor}
B_{\varphi} = \frac{B_{0}}{r\sin\theta}
\ee
with $B_{0}$ constant to the magnetic field. This term is curl-free,
divergence-free, and scales as the poloidal component of the magnetic
field.  Additionally, as $(\nabla\times\mathbf{B})\times\mathbf{B}$
is unchanged by this term, the equilibrium equations of the core
are independent of the toroidal component and, therefore, the
solutions for the density and flux functions obtained by Li \&
Shu~(\cite{ls96}) remain formally valid. We are aware that,
in general, one does not expect the toroidal component
of the field generated by the rotation of the core to be curl-free. 
Nevertheless, it is very instructive
to look at these idealised configurations with the aim of disentangling
column-density from magnetic-field effects in the propagation
of CR particles.

Substitution of Eq.~(\ref{FI}) into Eq.~(\ref{Bfield}), and the
addition of the toroidal component (Eq.~\ref{Btor}) results in
\be
\mathbf{B}(r,\theta) = \frac{4\pi c_s^2 r}{G^{1/2}}%
	\left[\frac{\ud\phi}{\ud\theta}\,\mathbf{\hat e}_r-%
	\phi\,\mathbf{\hat e}_\theta+b_0\phi(\pi/2)%
	\,\mathbf{\hat e}_\varphi\right]\,,
\ee
where 
\be
b_0 \equiv \frac{B_\varphi (r,\pi/2)}{B_{p}(r,\pi/2)}
\ee
is the ratio between the strength of the toroidal, $B_{\varphi}$,
and the poloidal, $B_{p}=(B_r^2+B_\theta^2)^{1/2}$, field in the
cloud's midplane. Notice that the magnetic field diverges on the
polar axis. However, this is not a problem for our calculations,
as the polar axis contains no mass.

\begin{figure}[!htp]
\begin{center}
\includegraphics[width=.45\textwidth, clip=true]{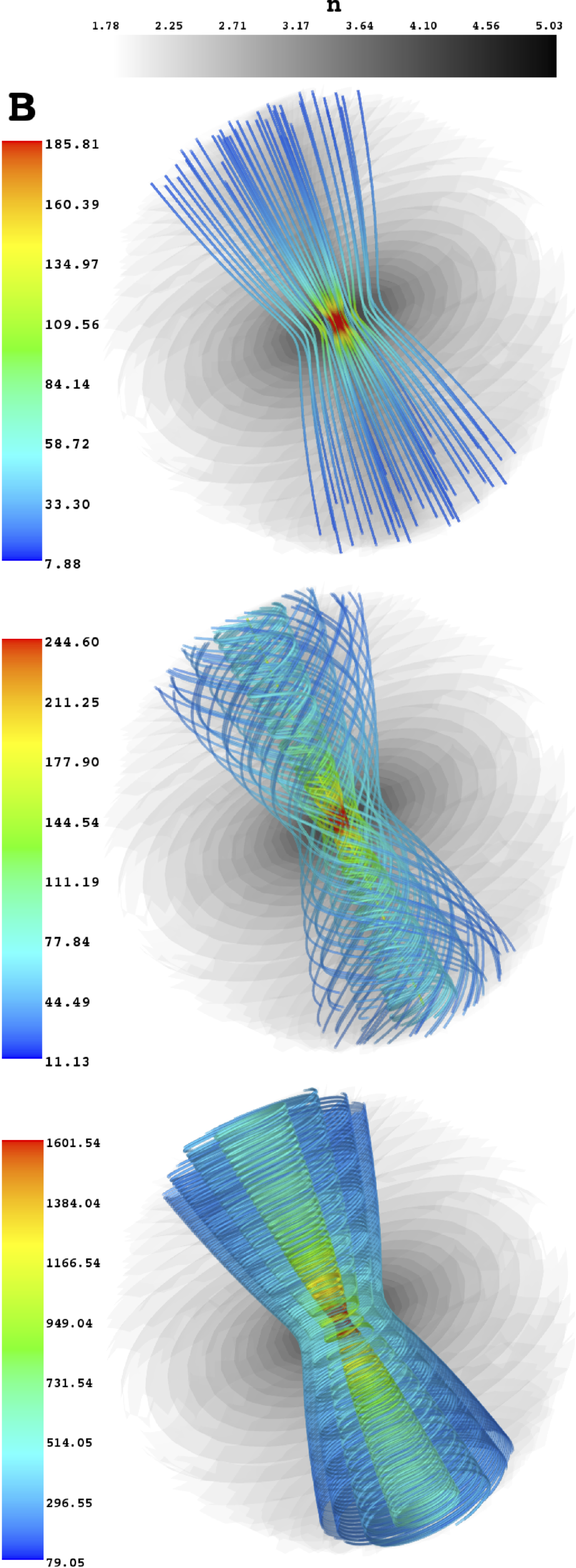}
\caption[]{Magnetic field lines of the $\lambda=2.66$ toroid with
$b_{0}=0$ (purely poloidal, \textit{upper plot}), $b_{0}=1$
(\textit{middle plot}), and $b_{0}=10$ (\textit{lower plot}). Iso-density
contours are shown in grey scales in unit of cm$^{-3}$ in
logarithmic scale, while the magnetic field module is shown in
colour scale in unit of $\mu$G. Field line plots are generated using
the visualisation software package MAYAVI (Ramachandran \&
Varoquaux~\cite{rv11}).}
\label{lishutor}
\end{center}
\end{figure}

The field lines that cross the midplane at $r=R_0$ and $\varphi=\varphi_0$,
are then be given by
\be
\label{rR0}
\frac{r(\theta)}{R_{0}}=\frac{\phi(\pi/2)}{\phi(\theta)}
\ee
and
\be
\varphi(\theta)=b_{0}\phi(\pi/2)\int_{\pi/2}^{\theta}%
\frac{\ud\vartheta}{\phi(\vartheta)\sin\vartheta}+%
\varphi_{0}\,.
\ee
As shown by Fig.~\ref{lishutor}, the magnetic field lines lie on
the surfaces of nested flux tubes defined by the condition $\Phi={\rm
const.}$, and their twisting increases as the distance from the
axis of symmetry decreases.

\subsection{Column-density versus magnetic-field effects on cosmic-ray ionisation rate}
\label{CDvsMAG}

During the propagation of a CR, the ionisation rate decreases due
to two main factors: $(i)$ CRs lose energy, thus their capability
of ionising hydrogen molecules, because of the increasing column
density ``seen'' by the particles themselves (PGG09); $(ii)$ magnetic
mirroring reduces the ionisation rate more than magnetic focusing
amplifies it (PG11).  In order to distinguish the origin of the
variation in the ionisation rate during the propagation, we assume
several values for $\lambda$ and $b_{0}$ resulting in different
density profiles and magnetic configurations.  

We assume three different values for the mass-to-flux ratio
($\lambda=8.38, 2.66,\mathrm{\,and \,} 1.63$) that acts modifying
the density distribution as well as the pinching of the magnetic
field lines.  $\lambda=8.38$ corresponds to the case of a ``roundish''
core, with a density distribution almost spherically symmetric;
when $\lambda=1.63$ the density is distributed in a disc-like
configuration; finally, $\lambda=2.66$ represents the intermediate
case. The pinching of magnetic field lines increases for decreasing
$\lambda$.

\begin{figure}[!h]
\includegraphics[width=.5\textwidth, clip=true]{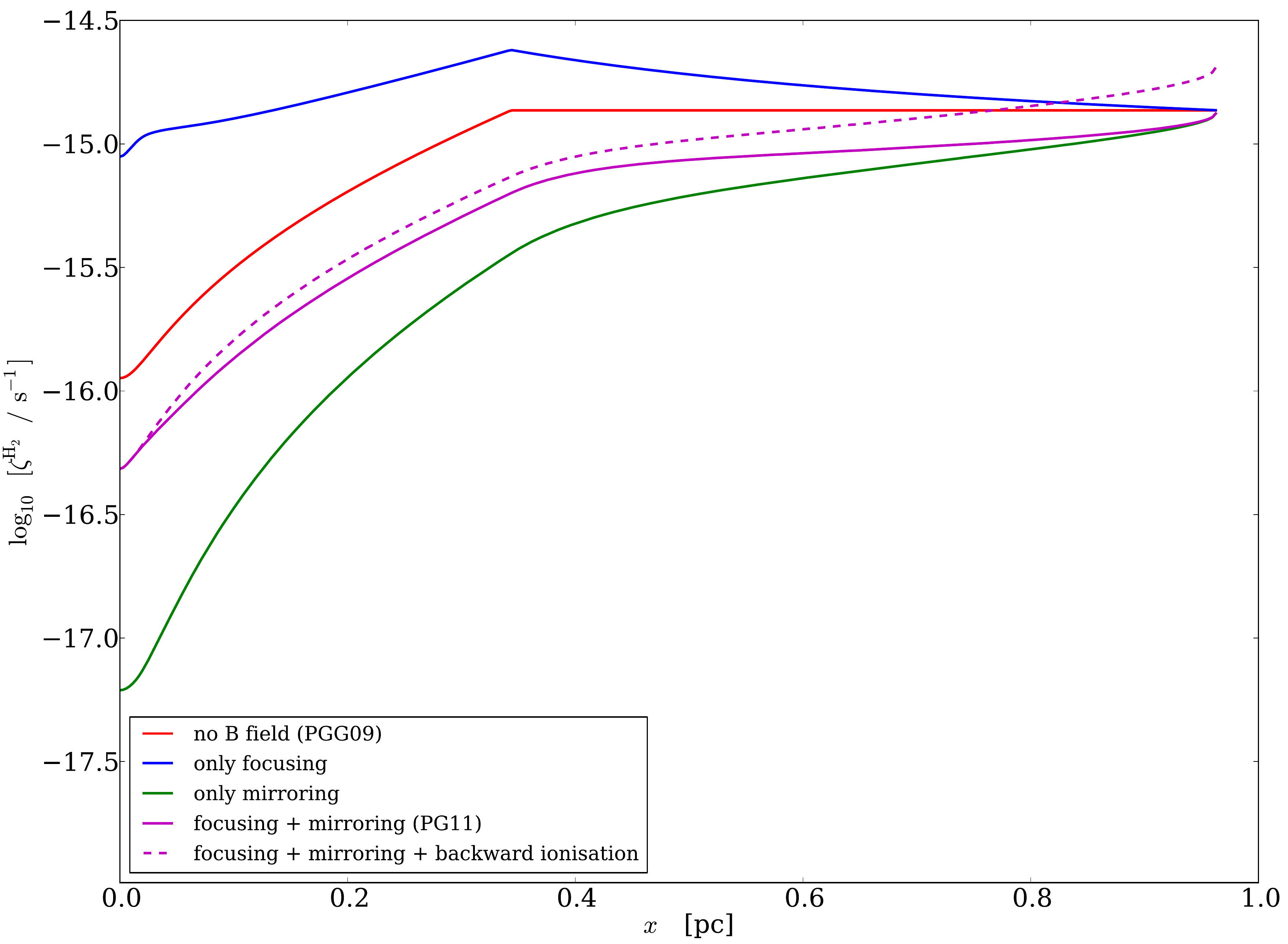}
\caption[]{Contributions to the ionisation rate as a function of the
position along the symmetry axis for the model $\cM$ and a flux
tube enclosing 1~$M_{\odot}$.}
\label{plot_paper_showbackion}
\end{figure}

We start assuming $b_{0}=0$ (purely poloidal field) whose effects
on CR propagation are described in PG11, then we increase the
toroidal field from $b_{0}=1$ (poloidal and toroidal fields have
comparable strengths) to $b_{0}=5,10, \mathrm{\,and\,} 50$.  We
follow the propagation of CRs entering the cloud from any direction
with all possible initial pitch angles applying the method described
in Sect.~\ref{method}.  With respect to PG11 we also consider the
fact that the mirrored CRs can still ionise while propagating
backwards. As expected, we find that this contribution to the
ionisation is stronger in the outer part of the cloud, namely in
the region also crossed by CRs which propagate backwards. This
further ionisation is not crucial since these CRs have already
passed through a large column density, losing their ionising power,
but now it is included in the model. Figure~\ref{plot_paper_showbackion}
shows the order of magnitude of this contribution for the model
$\cM$: we plot the trend of the CR ionisation rate for a flux tube
enclosing a mass of $1~M_{\odot}$ including all the contributions
deriving from mirroring, focusing, and backward ionisation.

\vspace{.6cm}

\subsection{Dependence of $\zeta^{\rm H_2}$ on the magnetic
field configuration}

\begin{figure*}[ht]
\centering
\includegraphics[width=17cm]{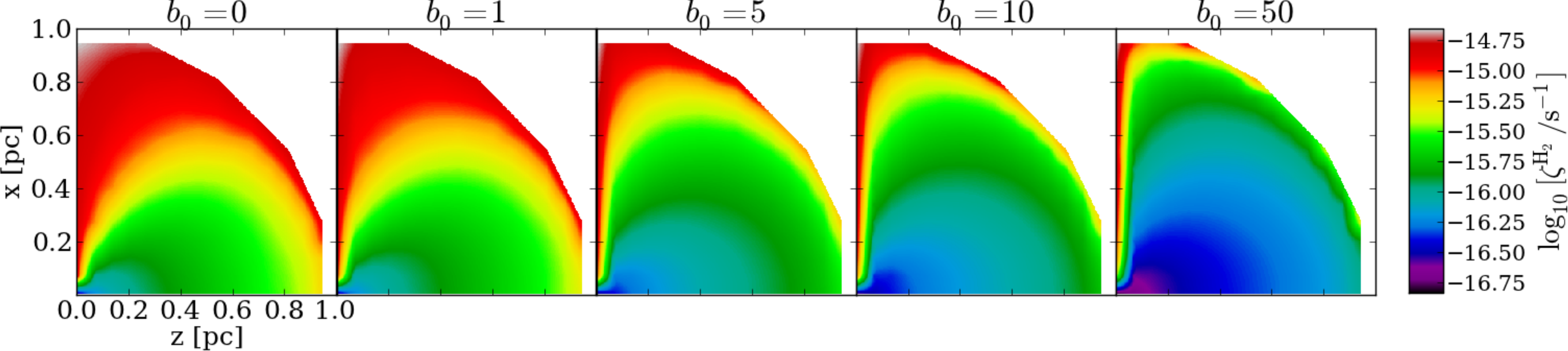}
\caption[]{CR ionisation rate profiles for the model $\cM$ in the plane crossing the symmetry
axis and perpendicular to the midplane $(y=0)$. The mass-to-flux
ratio is $\lambda=2.66$ and the strength of the toroidal field $b_{0}$ increases
from left to right.} 
\label{LiShuTor_H0_0.5_1pc_PAPER}
\end{figure*}

In order to investigate on the effects of a given magnetic field 
configuration on the variations of $\zeta^{\rm H_{2}}$, we assume a  
mass-to-flux ratio $\lambda=2.66$ so that the density shape is
fixed  
and we increase the magnitude
of the toroidal component (Fig.~\ref{lishutor}).
We find that increasing values of $b_{0}$ correspond to a decrease of 
$\zeta^{\rm H_2}$, in fact the area with low ionisation rate becomes
larger and larger reflecting the signatures of
the magnetic field configuration (Fig.~\ref{LiShuTor_H0_0.5_1pc_PAPER}).
Since $B\propto(r\sin\theta)^{-1}$, the intensity of the magnetic
field increases towards the symmetry axis.  Therefore, the focusing
factor becomes larger closer to the cloud's axis, and this explains
the increase of $\zeta^{\rm H_{2}}$ in this region.
  
As $b_0$ increases, the pitch angle $\alpha$ gets closer to $\pi/2$
quickly and the mirroring becomes more and more effective. However,
as we know from PG11, increasing the toroidal component of $B$ does
not change significantly the relative importance of mirroring versus
focusing effects, but it increases the effective column density as
seen by a CR gyrating around the field. If this increase in
$N(\mathrm{H}_{2})$ reaches the regime of exponential attenuation
($N(\mathrm{H}_2) \gtrsim 10^{25}$~cm$^{-2}$, see
Fig.~\ref{modelfitsAMRpaper}), the effect can be dramatic.  However,
this hardly happens in our semi-analytical model, which is taken
as representative of a pc-scale clump of modest column density.
Using profiles of density and magnetic field strength valid for
smaller spatial scales taken from simulations of collapsing clouds,
we will show that a reduction of 
$\zeta^{\rm H_{2}}$ of orders of magnitude can be achieved (see Sect.~\ref{nummod}).

\begin{figure}[]
\centering
\includegraphics[width=.5\textwidth, clip=true]{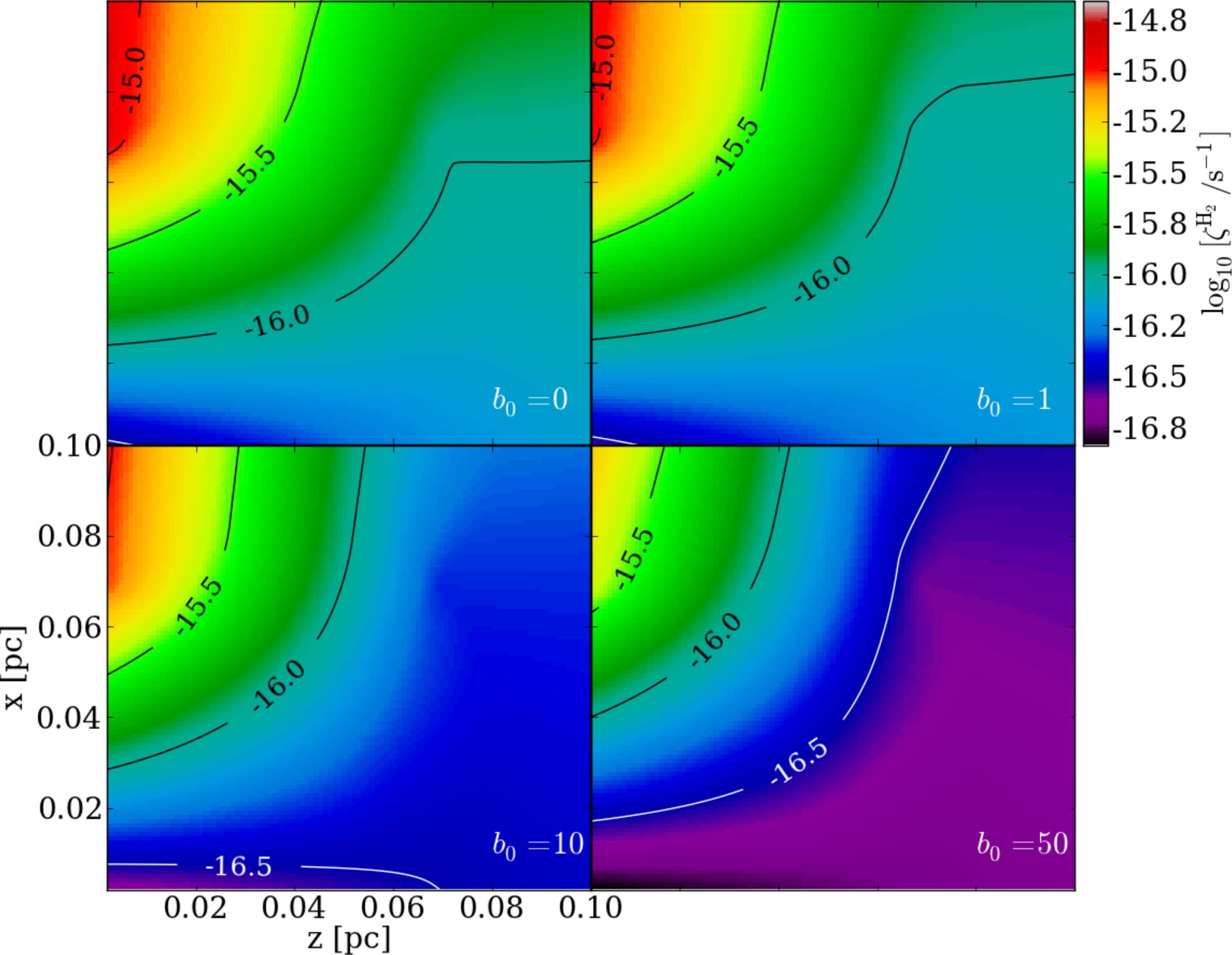}
\caption[]{CR ionisation rate maps for the model $\cM$ 
in the central 0.1~pc region in the plane
$y=0$ for $\lambda=2.66$ and increasing $b_{0}$. {\em Black} and
{\em white solid lines} show the iso-ionisation rate contours.}
\label{LiShuTor_zoom_zetaM}
\end{figure}

Figure~\ref{LiShuTor_zoom_zetaM} shows a zoom in the central region
at the core-size scale of 0.1~pc radius for the model $\cM$ in order
to appreciate more explicitly the reduction in the ionisation rate:
when the toroidal field is 50 times stronger than the poloidal field
component, a value of $\zeta^{\rm H_{2}}$ lower than the canonical
value of $3\times10^{-17}$~s$^{-1}$ ($\approx10^{-16.5}$~s$^{-1}$)
is readily reached in the region close to the cloud's midplane.

\subsection{Dependence of $\zeta^{\rm H_{2}}$ on the density profile}
\label{depondens}

\begin{figure}[h]
\center
\includegraphics[width=.4\textwidth, clip=true]{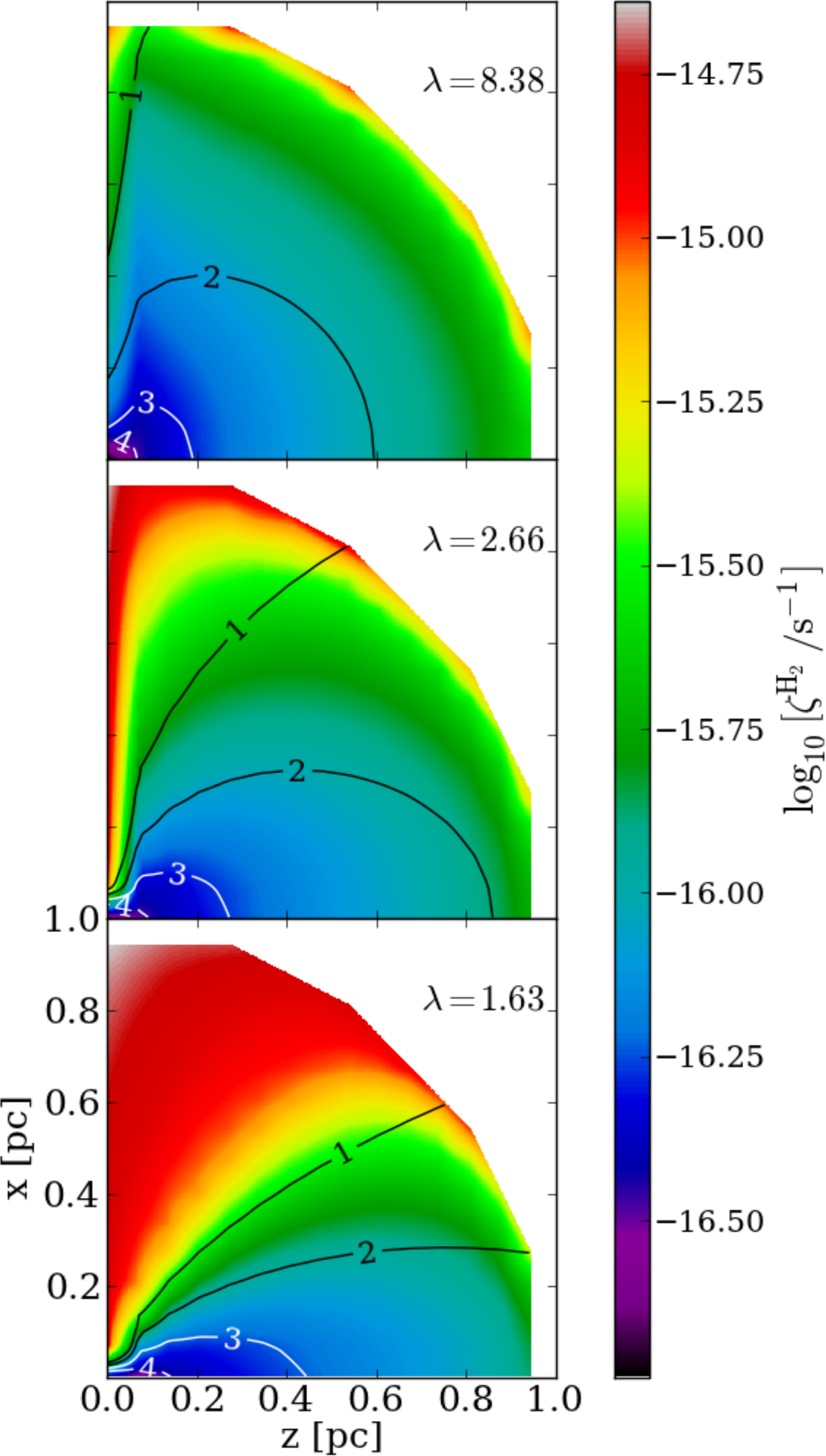}
\caption[]{CR ionisation rate maps for the model $\cM$ in the plane
$y=0$ for a fixed toroidal-to-poloidal ratio $b_{0}=10$ and different
values of the mass-to-flux-ratio.  {\em White and black contours}
represent the iso-density contours and the labels show $\log_{10}\
[n/\mathrm{cm^{-3}}]$.}
\label{LiShuTor_b0_10_PAPER}
\end{figure}

To examine the effects of the density configuration on the ionisation
rate, we fix a value of the toroidal-to-poloidal ratio and we vary
the mass-to-flux-ratio obtaining density profiles that span from a
``roundish'' core to a disc-like configuration (see Sect.~\ref{CDvsMAG}).
The effects of the column density can be promptly recognised by
noticing that the ionisation rate profile follows the shape of the
density profile.  In Fig.~\ref{LiShuTor_b0_10_PAPER} we superpose
the iso-density contours to the ionisation rate maps.  The regions
where iso-density contours follow the ionisation rate profile reveal
column-density effects.  Conversely, the areas where the iso-density
contours depart from the spatial distribution of $\zeta^{\rm H_{2}}$
are representative of magnetic-field effects.  In fact, for high
values of $\lambda$ the column density crossed is larger and
$\zeta^{\rm H_{2}}$ decreases rapidly (see PGG09). Said in a different
way, in the $\lambda=8.38$ model a particle ``sees'' a larger amount
of $N(\mathrm{H}_{2})$ and it loses more energy than in the the
flatter $\lambda=1.63$ model where the density along the field
direction increases only close to the cloud's midplane.

\section{Numerical models}
\label{nummod}

In this Section we describe the ionisation rate maps obtained from
a number of numerical simulations related to a collapsing rotating
core performed with the AMR code RAMSES\footnote{RAMSES simulations
were analysed using PyMSES (Labadens et al.~\cite{lc11}).}
(Teyssier~\cite{t02}, Fromang et al.~\cite{fh06}) and detailed in
Joos et al.~(\cite{jh12}). They considered a spherical $1~M_{\odot}$
cloud with an initial density profile
\be
n(r)=\frac{n_0}{1+(r/r_0)^2}\,,
\ee\noindent
where $n_{0}=7.8\times 10^6$~cm$^{-3}$ 
and $r_0=4.68\times 10^{-3}$~pc according to observations
(Andr\'e et al.~\cite{awt00}, Belloche et al.~\cite{ba02}). The
thermal-to-gravitational energy ratio is about 0.25 and the
rotational-to-gravitational energy ratio is about 0.03. We select
a series of simulations from Joos et al.~(\cite{jh12}) varying the
mass-to-flux ratio $\lambda$ and the angle between the initial
magnetic field direction and the initial rotation axis
$\alpha_\mathrm{B,J}$.  Table~\ref{lambdaalfatime} lists the
parameters.

\begin{table}[!h]
\caption{Parameters of the simulations described in the text (from
Joos et al.~\cite{jh12}): mass-to-flux ratio, initial angle between
the magnetic field direction and the rotation axis, time after the
formation of the first Larson's core (core formed in the centre of
the pseudo-disc with $n\gtrsim10^{10}$~cm$^{-3}$ and $r\sim10-20$~AU),
maximum mass of the protostellar core and of the disc.  Last column
gives information about the disc formation.}
\begin{tabular}{ccccccc}
\hline
Case & $\lambda$ & $\alpha_{\rm B,J}$ & $t$ & $M_{\bigstar}$ & $M_{\rm disc}$ & Disc ?\\
 &         & [rad]                 & [kyr] & $[M_{\odot}]$ & $[M_{\odot}]$ & (Y$^{a}$ / N$^{b}$ / K$^{c}$)\\
\hline\hline
A$_{1}$ & 5  &  0  &  0.824 & -- & -- & N\\
A$_{2}$ & 5  &  0  &  11.025 & 0.26 & 0.05 & N\\
B & 5  &  $\pi$/4 & 7.949 & 0.23 & 0.15 & Y\\
C & 5  &  $\pi$/2 & 10.756 & 0.46 & 0.28 & K\\
D & 2  &  0  &  5.702 & 0.24 & -- & N\\
E & 17 &  0  &  6.620 & 0.43 & 0.15 & K\\
\hline
\end{tabular}\\[2pt]
\noindent $^a$ A disc with flat rotation curve is formed (Fig.~15 in Joos et al.~\cite{jh12}).\\
\noindent $^b$ No significant disc is formed ($M_{\rm disc}<5\times10^{-2}~M_{\odot}$).\\
\noindent $^c$ A keplerian disc is formed (Fig.~14 in Joos et al.~\cite{jh12}).

\label{lambdaalfatime}
\end{table}%

According to the method described in Sect.~\ref{method} we compute
the ionisation rate maps making use of the model $\cM$.
For each case in Table~\ref{lambdaalfatime}, we show the maps of
$\zeta^{\rm H_{2}}$ in a plane parallel and perpendicular to the
main direction of the magnetic field (always along the $x$ axis)
and containing the density peak,
namely the $(z,x)$ and the $(y,z)$ plane, respectively.  Besides,
in Figures~\ref{RAMSES_M_0.3_0_100}--\ref{RAMSES_M_0.1_0_44} we
plot the results for the whole computational domain (upper and
middle left plots) and zooming into the inner 1000~AU (upper and
middle right plots) plus a graph showing the magnetic field line
morphology in the central 600~AU (lower plot).

We are aware of the fact that the ionisation rate should be computed
simultaneously to the MHD model. This will be the subject of a
future work, but the results described in this paper can be considered
an important proof of concept that very low ionisation rate
can be achieved in the inner regions of a collapsing cloud.

\subsection{Intermediate magnetisation ($\lambda=5$)}
\label{intermediate}

We consider a couple of outputs (cases A$_{1}$ and A$_{2}$ in
Table~\ref{lambdaalfatime}) for an aligned rotator
($\alpha_\mathrm{B,J}=0$) and super-critical cloud ($\lambda=5$)
in agreement with observations (Crutcher~\cite{c99}).  We choose
two different times after the formation of the first Larson's core
in order to understand the effect of the tangling of magnetic field
lines on CR propagation when a disc is not formed.  Unlike the
semi-analytical case (Sect.~\ref{semi-analytical}) where the density
does not even achieve $10^{5}$~cm$^{-3}$ and the symmetry of the
magnetic field configuration is conserved for any toroidal-to-poloidal
ratio, in all the numerical models presented here the central density
reaches the value of about $10^{13}$~cm$^{-3}$ and in general it
is higher than $10^{10}$~cm$^{-3}$ in the inner $50-100$~AU radius.
Besides, the symmetry of magnetic field lines is broken very soon
with time.
This is likely due to the development of the interchange instability
(Krasnopolsky et al.~\cite{kl12}).

\begin{figure}[!h]
\centering
\subfigure{\includegraphics[width=.5\textwidth, clip=true]{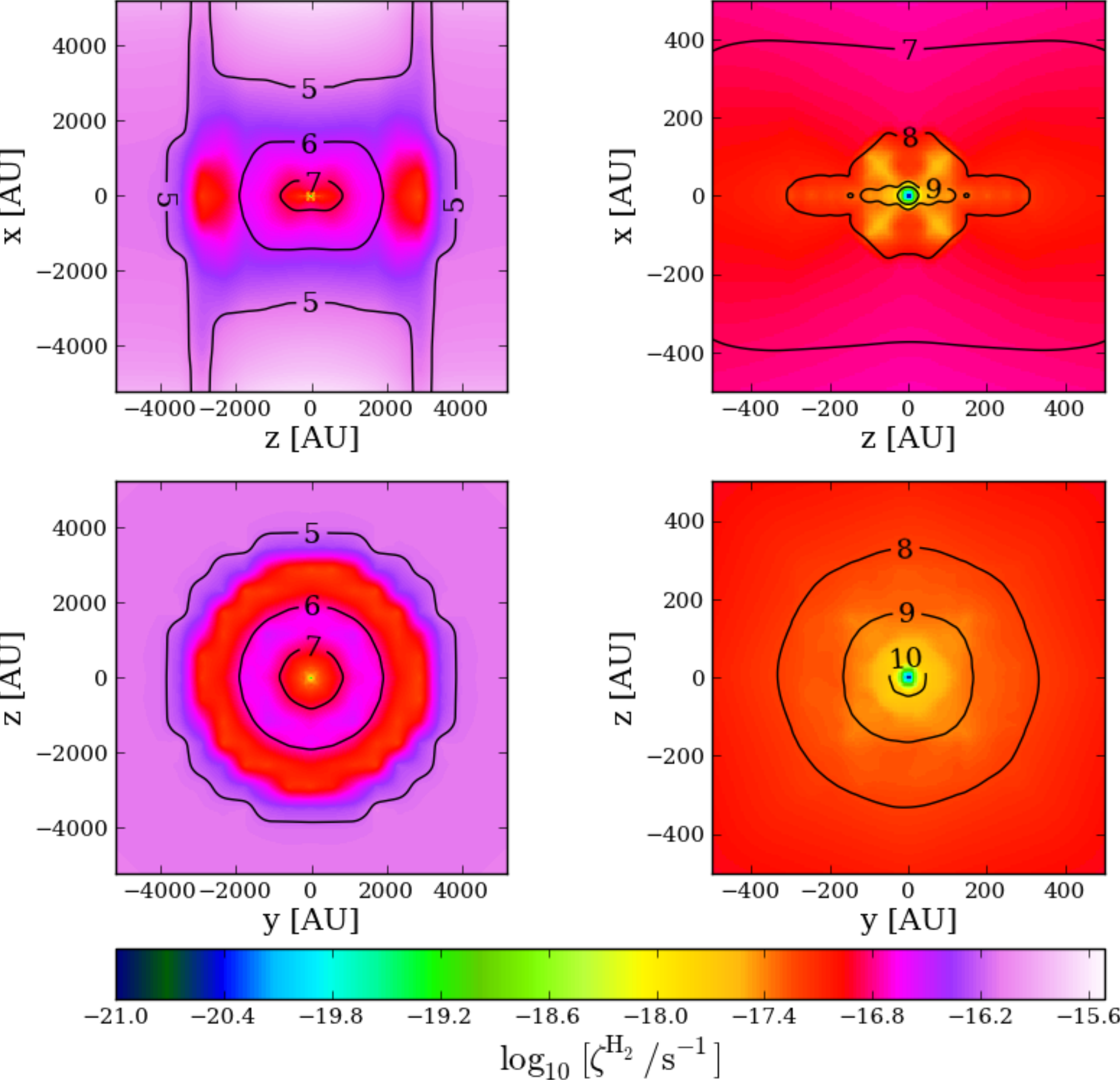}}
\subfigure{\includegraphics[width=.5\textwidth, clip=true]{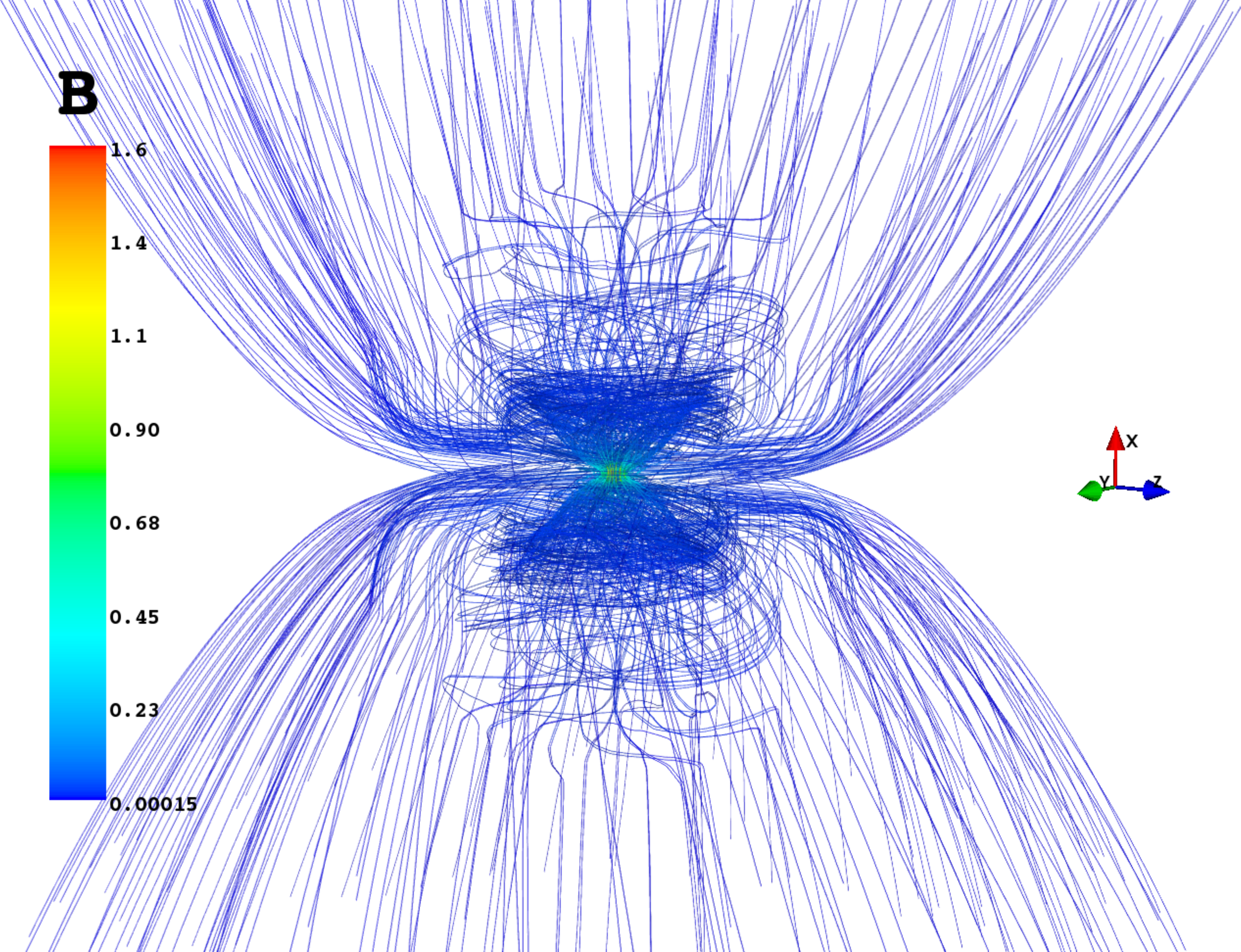}}
\caption[]{CR ionisation maps and iso-density contours ({\em black
solid lines}) for the case A$_{1}$ in Table~\ref{lambdaalfatime}.
{\em Upper and middle left} panels show the entire computational
domain while {\em upper and middle right} panels show a zoom in the
inner region.  {\em Upper} panels show a slice of a plane parallel
to the magnetic field direction, while {\em middle} panels refer
to a slice of a perpendicular plane.  Both planes contain the density
peak and labels show $\log_{10}\ [n/\mathrm{cm^{-3}}]$.  The {\em
lower} plot shows the magnetic field line morphology in the central
600~AU of the RAMSES data cube. The colour bar shows the magnetic
field module in Gauss units.}
\label{RAMSES_M_0.3_0_100}
\end{figure}

We know from the semi-analytical model (Sect.~\ref{semi-analytical})
that it is not possible to disentangle column-density from magnetic
effects, but both intervene on the decrease of the ionisation rate.
In Section~\ref{fittingformula} we give an estimate of the relative
incidence of these two effects.  As explained in Sect.~\ref{depondens},
we can interpret the deviations between the iso-density contours
and $\zeta^{\rm H_{2}}$ maps as due to magnetic imprints. As an
instance, in the upper right plot of Fig.~\ref{RAMSES_M_0.3_0_100}
there is a clear departure between the contours at $n=10^{9}$~cm$^{-3}$
and the shape of the region where $\zeta^{\rm H_{2}}$ reaches values
of $2-3\times10^{-18}$~s$^{-1}$ (in yellow in the Figure).  Another
example is the upper left plot of the same Figure where the region
in red with $\zeta^{\rm H_{2}}\sim1-2\times10^{-17}$~s$^{-1}$ extends
for densities spanning from $10^{5}$ to more than $10^{7}$~cm$^{-3}$
in a horizontal ``strip'' of about 5000~AU along the $z$ axis. This
calls to mind the magnetic field configuration (see lower panel of
Fig.~\ref{RAMSES_M_0.3_0_100}), in fact this is the region where
field lines start to be twisted due to rotation.  
The middle right panel of Fig.~\ref{RAMSES_M_0.3_0_100} shows that less
than $10^{3}$~yr after the formation of the first Larson's core, a
central region with $r\sim100-200$~AU is characterised by $\zeta^{\rm
H_{2}}\sim2-4\times10^{-18}$~s$^{-1}$.

\begin{figure}[!h]
\centering
\subfigure{\includegraphics[width=.5\textwidth, clip=true]{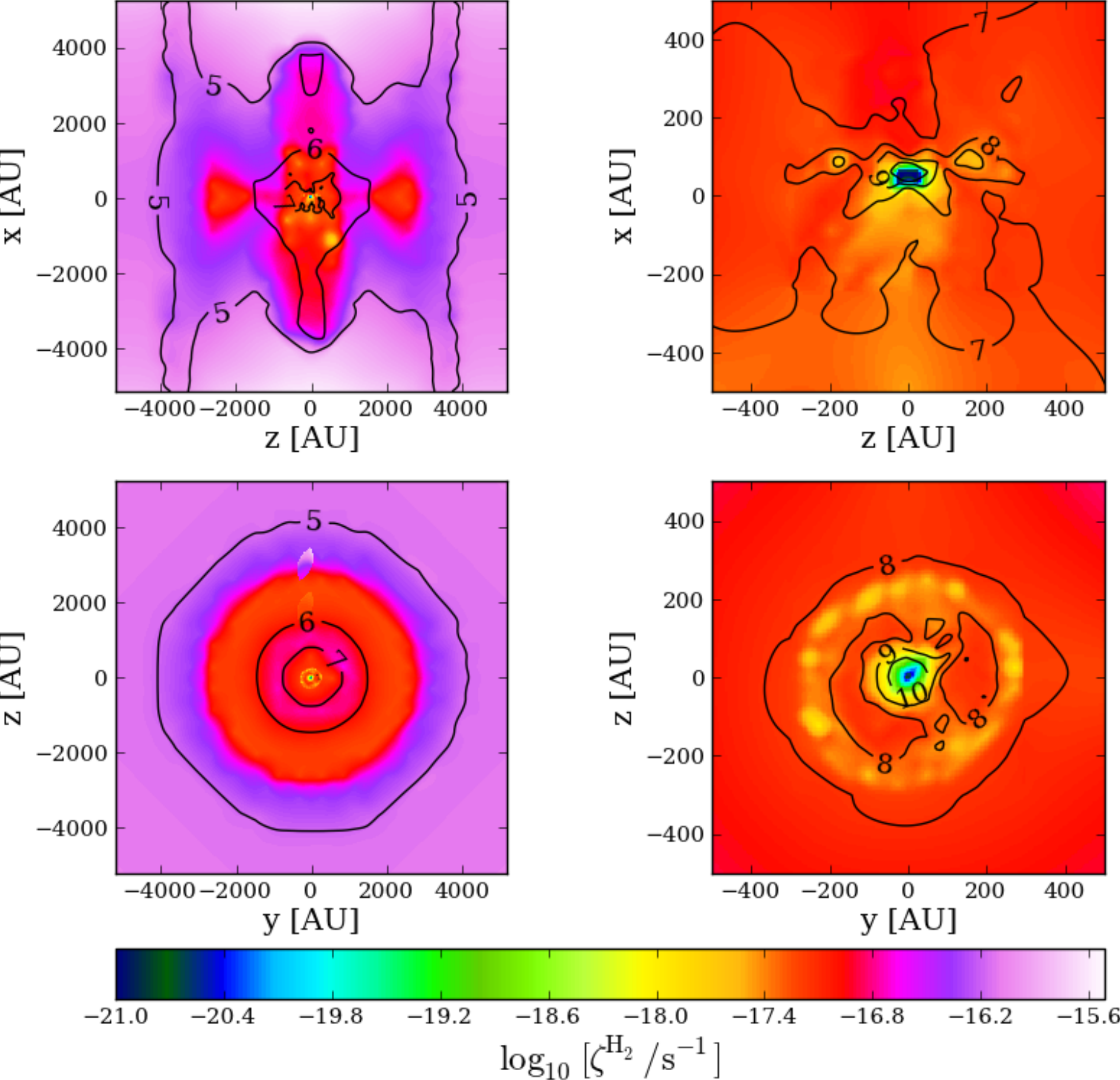}}
\subfigure{\includegraphics[width=.48\textwidth, clip=true]{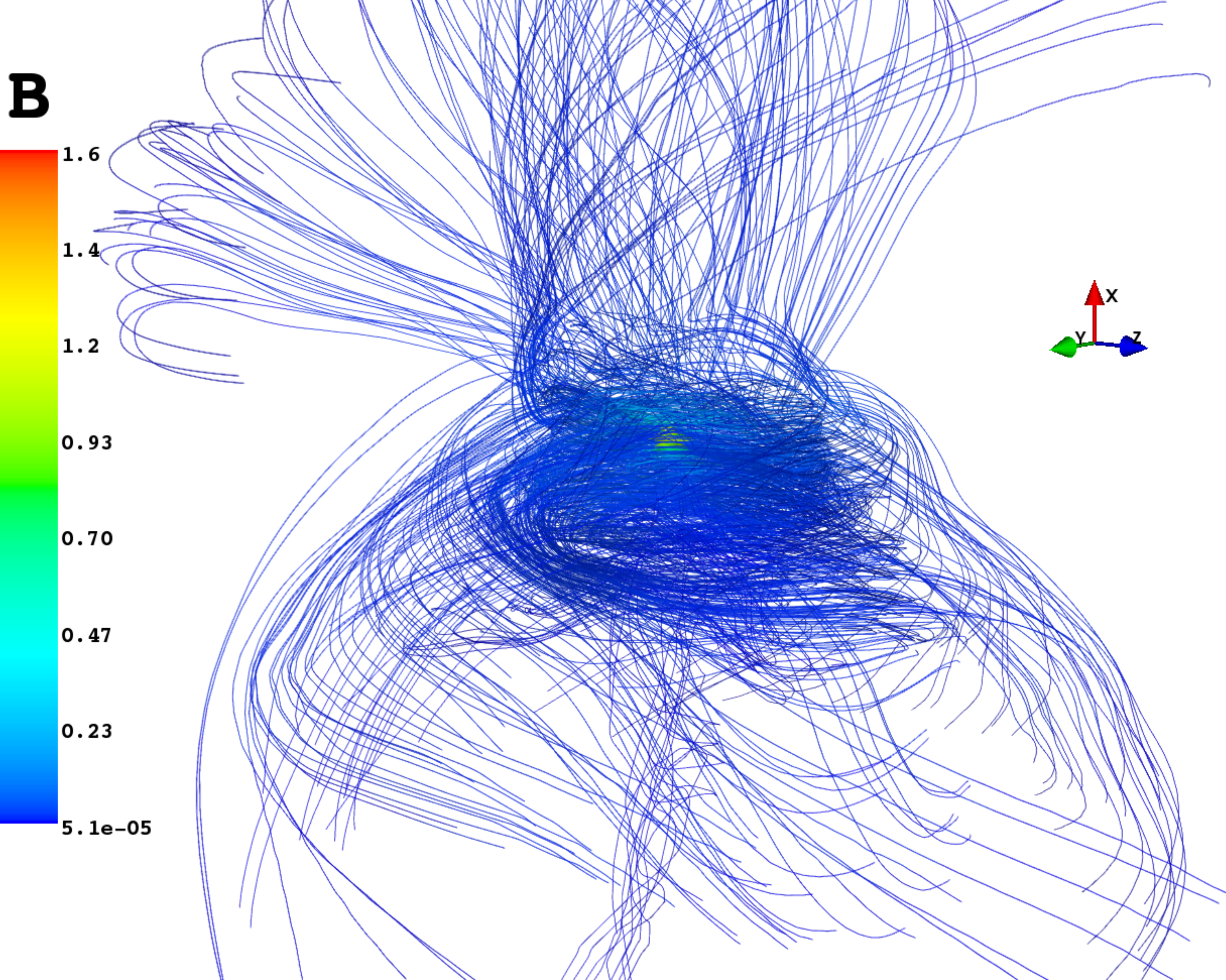}}
\caption[]{CR ionisation maps and iso-density contours ({\em black solid
lines}) for the case A$_{2}$ in Table~\ref{lambdaalfatime}. See
Fig.~\ref{RAMSES_M_0.3_0_100} for further information.}
\label{RAMSES_M_0.3_0_350}
\end{figure}

We compute $\zeta^{\rm H_{2}}$ for a later-time configuration (case
A$_{2}$) with the same initial conditions ($\lambda=5$,
$\alpha_\mathrm{B,J}=0$). As previously mentioned, we lose the
symmetry of field lines (see lower panel of Fig.~\ref{RAMSES_M_0.3_0_350})
as well as of the density profile (see upper and middle right panels
of Fig.~\ref{RAMSES_M_0.3_0_350}).  At large scales (upper and
middle left plots), we notice that the region with $\zeta^{\rm
H_{2}}\lesssim10^{-17}$~s$^{-1}$ is less extended along the $z$
axis and it is elongated parallel to the $x$ axis (i.e., parallel
to the magnetic field).  At small scales (upper and middle right
plots), a flattened structure formed almost perpendicularly to the
rotation axis can be noticed.  In fact, even if the disc is not
formed, the plane perpendicular to the rotation axis shows the
presence of a ``ring'' with densities between 10$^{8}$ and
10$^{9}$~cm$^{-3}$ and average ionisation rate of about 10$^{-18}$~s$^{-1}$
circumscribing the density peak up to a radius of about 300~AU.  

\begin{figure}[!h]
\centering
\subfigure{\includegraphics[width=.5\textwidth, clip=true]{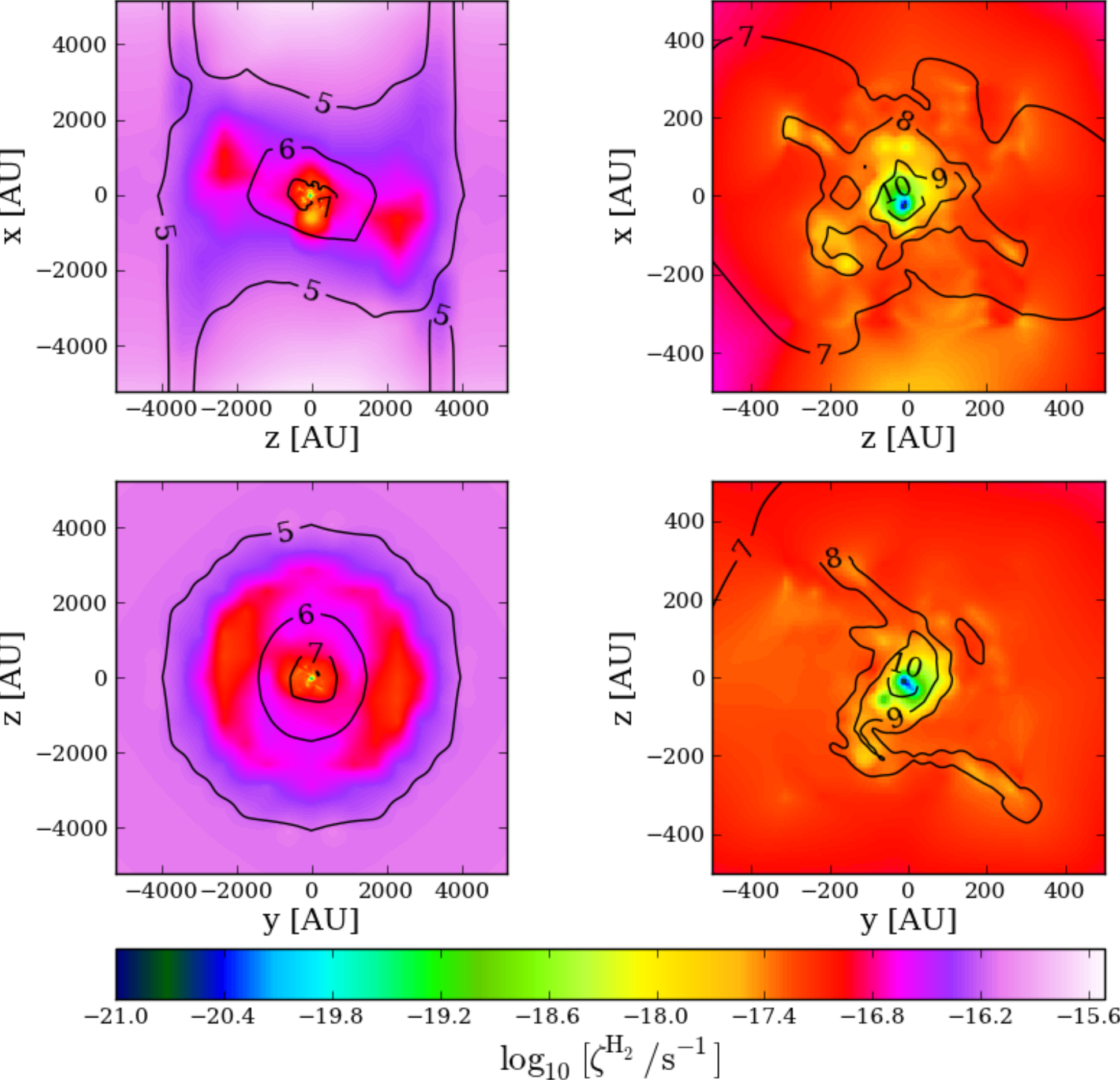}}
\subfigure{\includegraphics[width=.48\textwidth, clip=true]{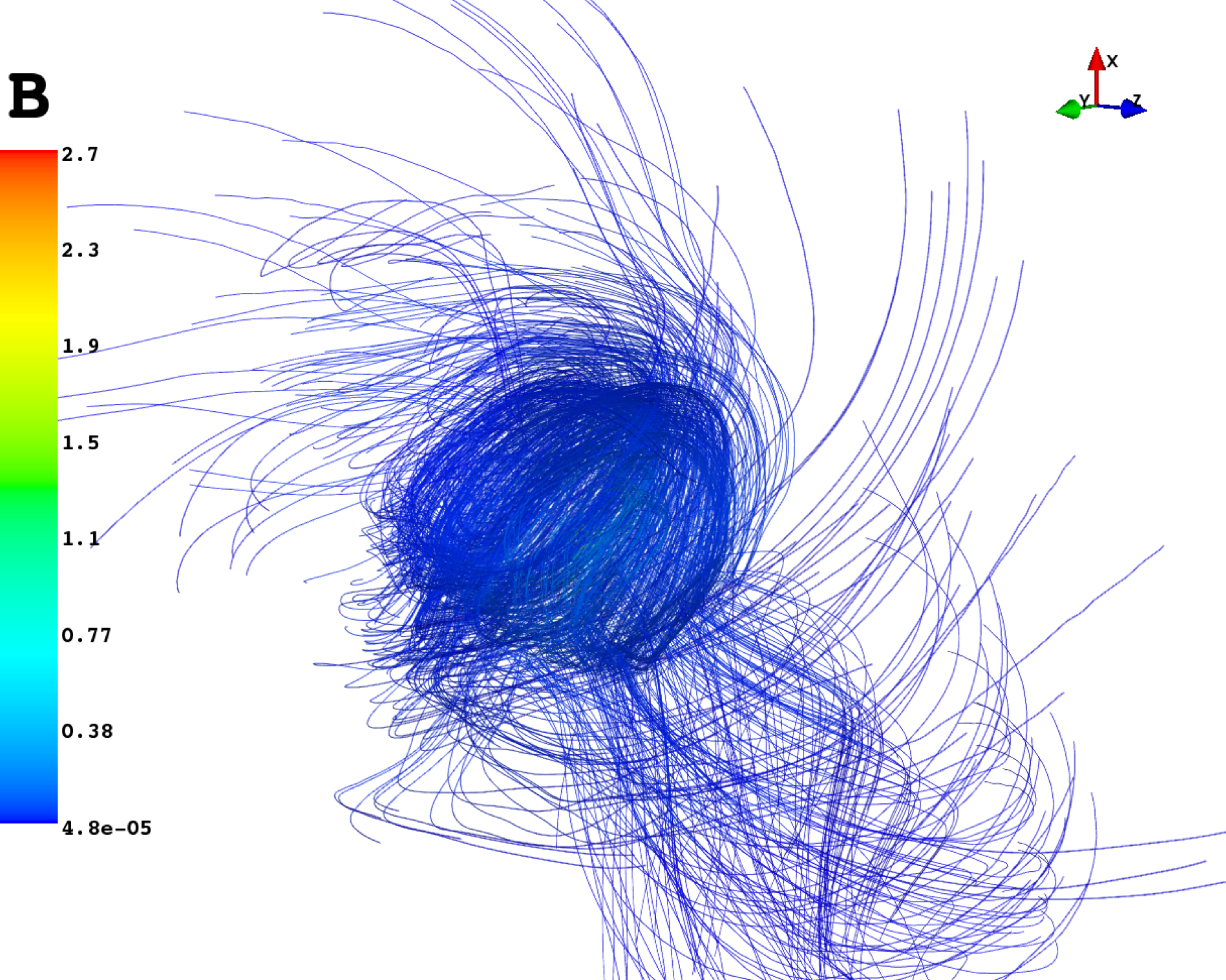}}
\caption[]{CR ionisation maps and iso-density contours ({\em black solid
lines}) for the case B in Table~\ref{lambdaalfatime}. See
Fig.~\ref{RAMSES_M_0.3_0_100} for further information.}
\label{RAMSES_M_0.3_0.8_58}
\end{figure}

We account for a configuration with the initial rotation axis twisted
by $\pi/4$ towards the $y$ axis, while the initial magnetic field
direction is still along the $x$ axis (case B in
Table~\ref{lambdaalfatime}).  This case predicts the formation of
a disc with a flat rotation curve in the plane perpendicular to the
rotation axis. In the central $r\sim50$~AU region of both the $(z,x)$
and the $(y,z)$ plane, we find a CR ionisation rate lower than about $10^{-19}$~s$^{-1}$.
In the inner 500~AU region (upper and middle right panels
of Fig.~\ref{RAMSES_M_0.3_0.8_58}) we can still identify a relationship
between $\zeta^{\rm H_{2}}$ and $n$, even if the density profile
becomes more and more irregular and field lines almost lose memory
of their initial configuration (see lower panel of
Fig.~\ref{RAMSES_M_0.3_0.8_58}).

\begin{figure}[!h]
\centering
\subfigure{\includegraphics[width=.5\textwidth, clip=true]{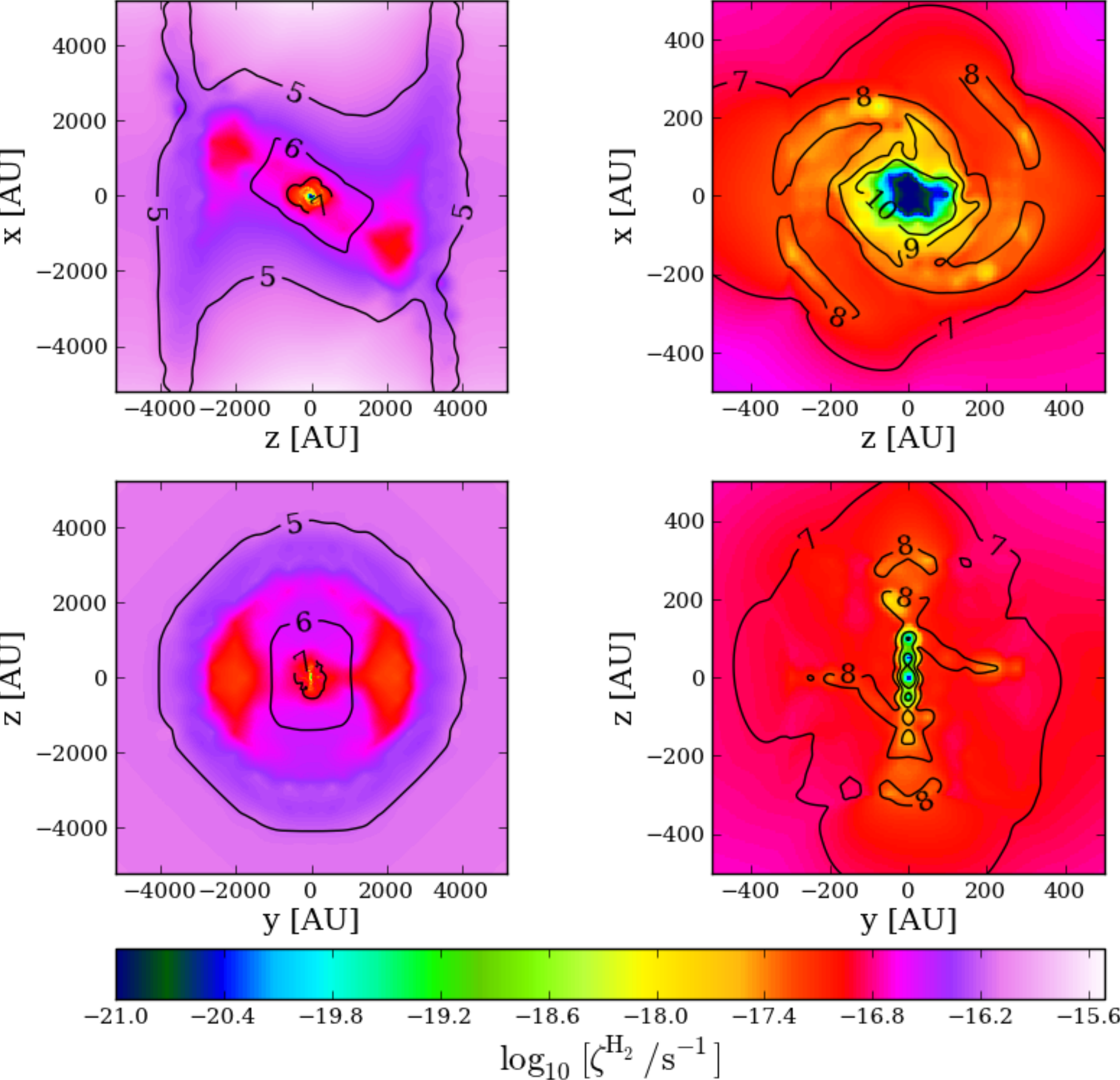}}
\subfigure{\includegraphics[width=.505\textwidth, clip=true]{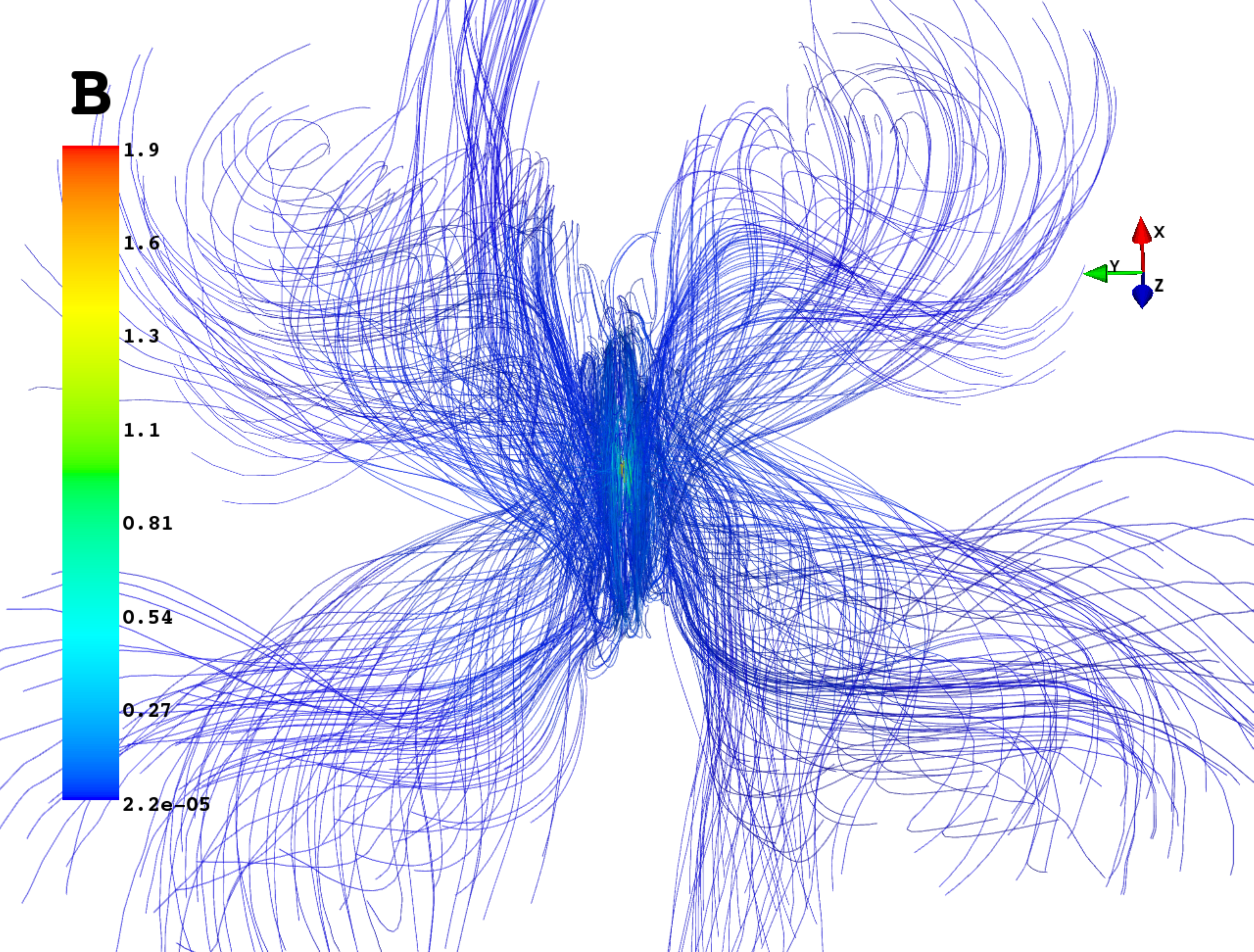}}
\caption[]{CR ionisation maps and iso-density contours ({\em black solid
lines}) for the case C in Table~\ref{lambdaalfatime}. See
Fig.~\ref{RAMSES_M_0.3_0_100} for further information.}
\label{RAMSES_M_0.3_1.6_187}
\end{figure}

Finally, we consider the case of a perpendicular rotator
($\alpha_\mathrm{B,J}=\pi/2$) so that initially the rotation axis
is along the $y$ axis. In this circumstance (case C in
Table~\ref{lambdaalfatime}), a keplerian disc perpendicular to the
rotation axis is predicted. Figure~\ref{RAMSES_M_0.3_1.6_187} shows
this configuration, namely a face-on view in the $(z,x)$ plane and
an edge-on view of the disc in the $(y,z)$ plane.  It is worth
noting in the face-on view (upper right panel of
Fig.~\ref{RAMSES_M_0.3_1.6_187}) a large region of $r\sim200$~AU
and $n\gtrsim10^{9}$~cm$^{-3}$ with $\zeta^{\rm
H_{2}}\lesssim10^{-18}$~s$^{-1}$. Here the ratio between the toroidal
component and the total magnetic field, $B_{\varphi}/|{\bf B}|$,
is larger than about 0.4. We reach even lower values, with a minimum
of $2\times10^{-21}$~s$^{-1}$ in the inner area that has an extent
of a few tens of AU. 
The CR ionisation rate is so low that we can assume the gas to be
effectively decoupled with the magnetic field. A similar behaviour
can be also appreciated in the edge-on
view (lower right panel of Fig.~\ref{RAMSES_M_0.3_1.6_187}). 
The very low value of $\zeta^\mathrm{H_{2}}$ found in this collapse
region corresponds to the condition required by Mellon \& Li~(\cite{ml09}) for the
formation of a 10~AU disc when $\lambda=4$.
However, we stress that in this case the formation of the disc is made possible
by the misalignment of the angular momentum and the magnetic field of the 
cloud, whereas in the situation considered by Mellon \& Li~(\cite{ml09}) the two
vectors are aligned and disc formation is enabled by the enhanced ambipolar diffusion
resulting from the lower value of $\zeta^{{\rm H}_2}$.
Finally, the lower panel of
Fig.~\ref{RAMSES_M_0.3_1.6_187} shows how the rotation perpendicular
to the $y$ axis forces the field lines (initially with a poloidal
configuration along the $x$ axis) to wrap around the rotation axis.

\subsection{Strong magnetisation ($\lambda=2$)}

\begin{figure}[!h]
\centering
\subfigure{\includegraphics[width=.5\textwidth, clip=true]{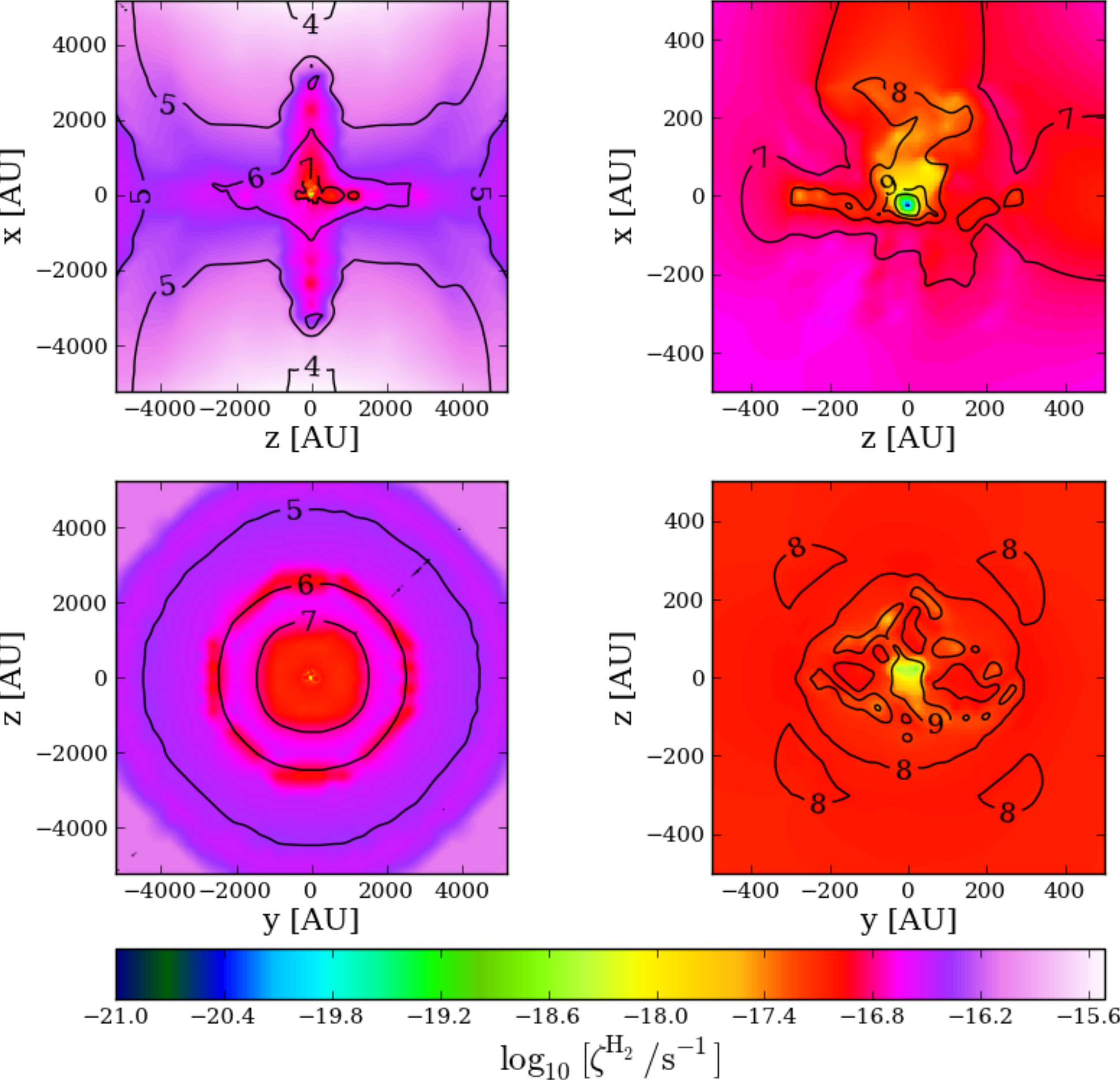}}
\subfigure{\includegraphics[width=.5\textwidth, clip=true]{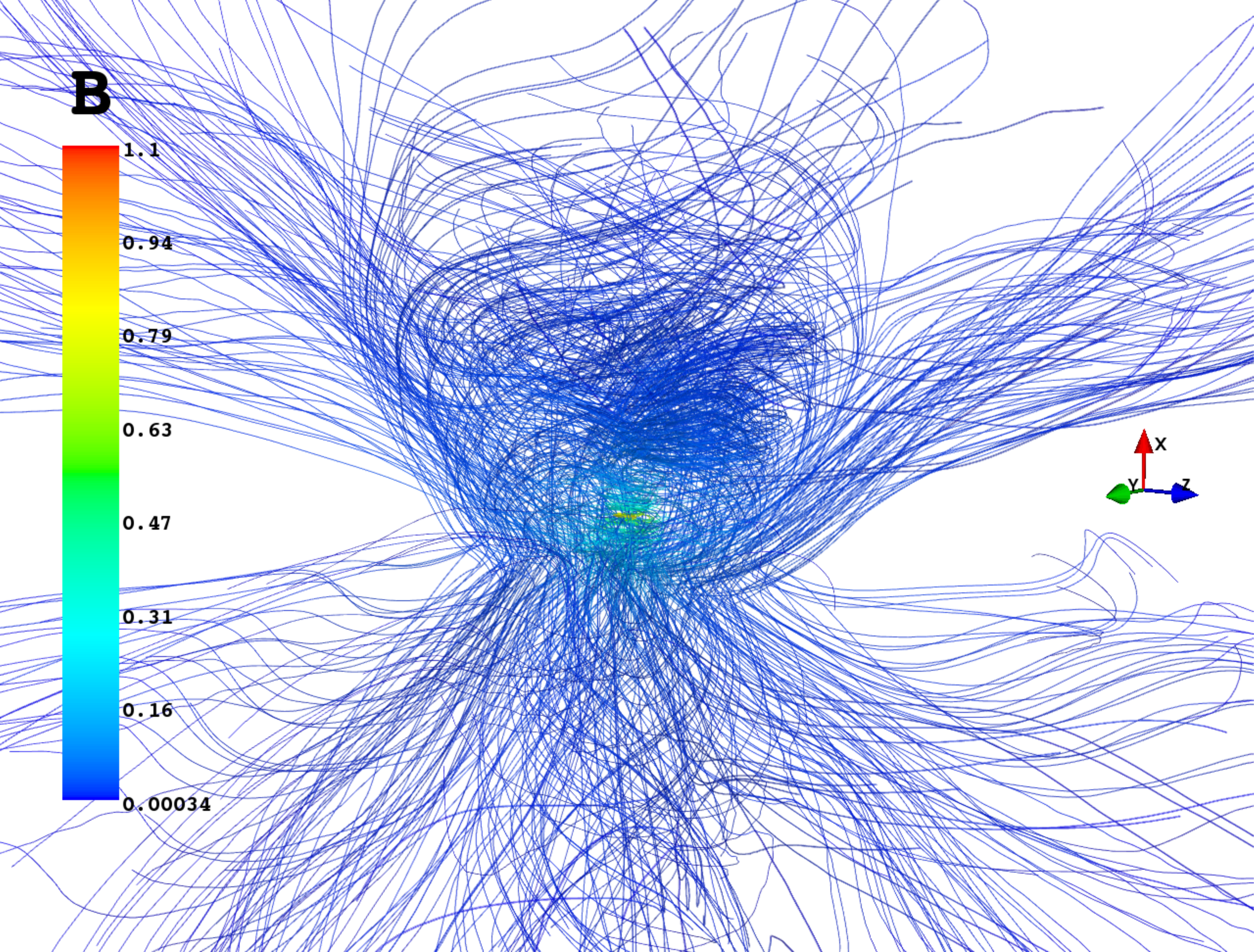}}
\caption[]{CR ionisation maps and iso-density contours ({\em black solid
lines}) for the case D in Table~\ref{lambdaalfatime}. See
Fig.~\ref{RAMSES_M_0.3_0_100} for further information.}
\label{RAMSES_M_0.7_0_87}
\end{figure}

We analyse the case of an aligned rotator with strong magnetic field
for a late-time configuration (case D of Table~\ref{lambdaalfatime}).
A stronger field is more resistant to line twisting caused by
rotation. In fact, in this case, the poloidal configuration can be
still identified after about $6$~kyr from the formation of the
first Larson's core (see lower panel of Fig.~\ref{RAMSES_M_0.7_0_87}).
As expected, the effect of magnetic braking is remarkable and the
disc is not formed in the plane perpendicular to the rotation axis
(see middle right panel of Fig.~\ref{RAMSES_M_0.7_0_87}). In this
case we find the ionisation rate to be more uniform and close to
$10^{-17}$~s$^{-1}$, while in the inner $50-100$~AU $\zeta^{\rm
H_{2}}$ is of the order of $10^{-19}$~s$^{-1}$ in the $(z,x)$ plane
and $10^{-18}$~s$^{-1}$ in the $(y,z)$ plane.  Density contours in
the $(y,z)$ plane reveal high density fluctuations not allowing the
formation of large high-density structures.

Since most molecular cloud cores appear to have significant levels of
magnetisation (Crutcher~1999), this case is probably the most relevant 
for modelling cloud collapse. Our findings that the CR ionisation rate is 
one-two orders of magnitudes below the interstellar value in a region around the 
accreting protostar of radius $\sim 50$--$100$~AU, could indicate
that a centrifugally supported disc of this size might form even in the strongly 
magnetised case. Of course,
this question can only be answered by performing a fully self-consistent 
dynamical calculation of CR propagation during cloud collapse.

\subsection{Weak magnetisation ($\lambda=17$)}

\begin{figure}[!t]
\centering
\subfigure{\includegraphics[width=.5\textwidth, clip=true]{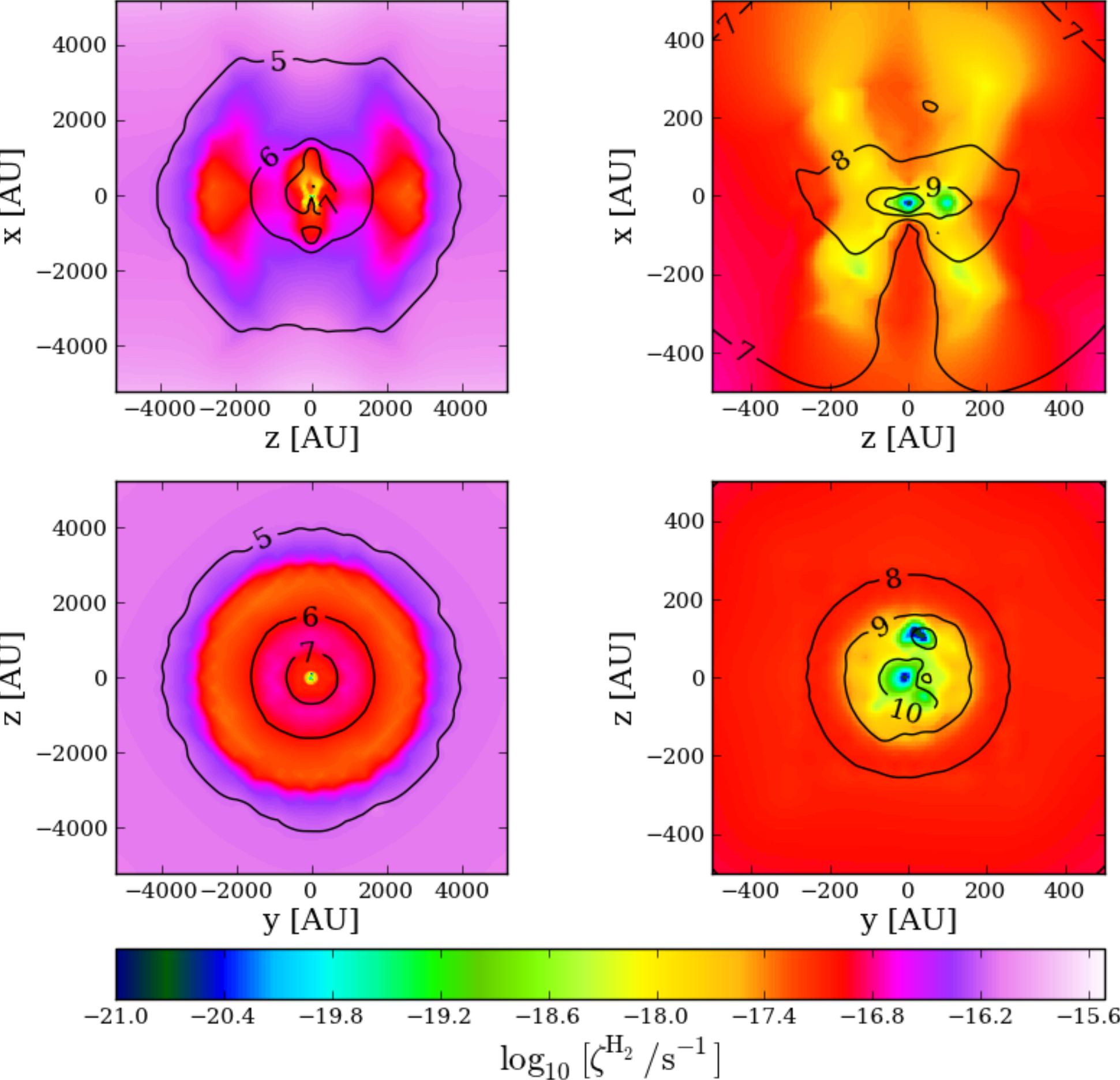}}
\subfigure{\includegraphics[width=.38\textwidth, clip=true]{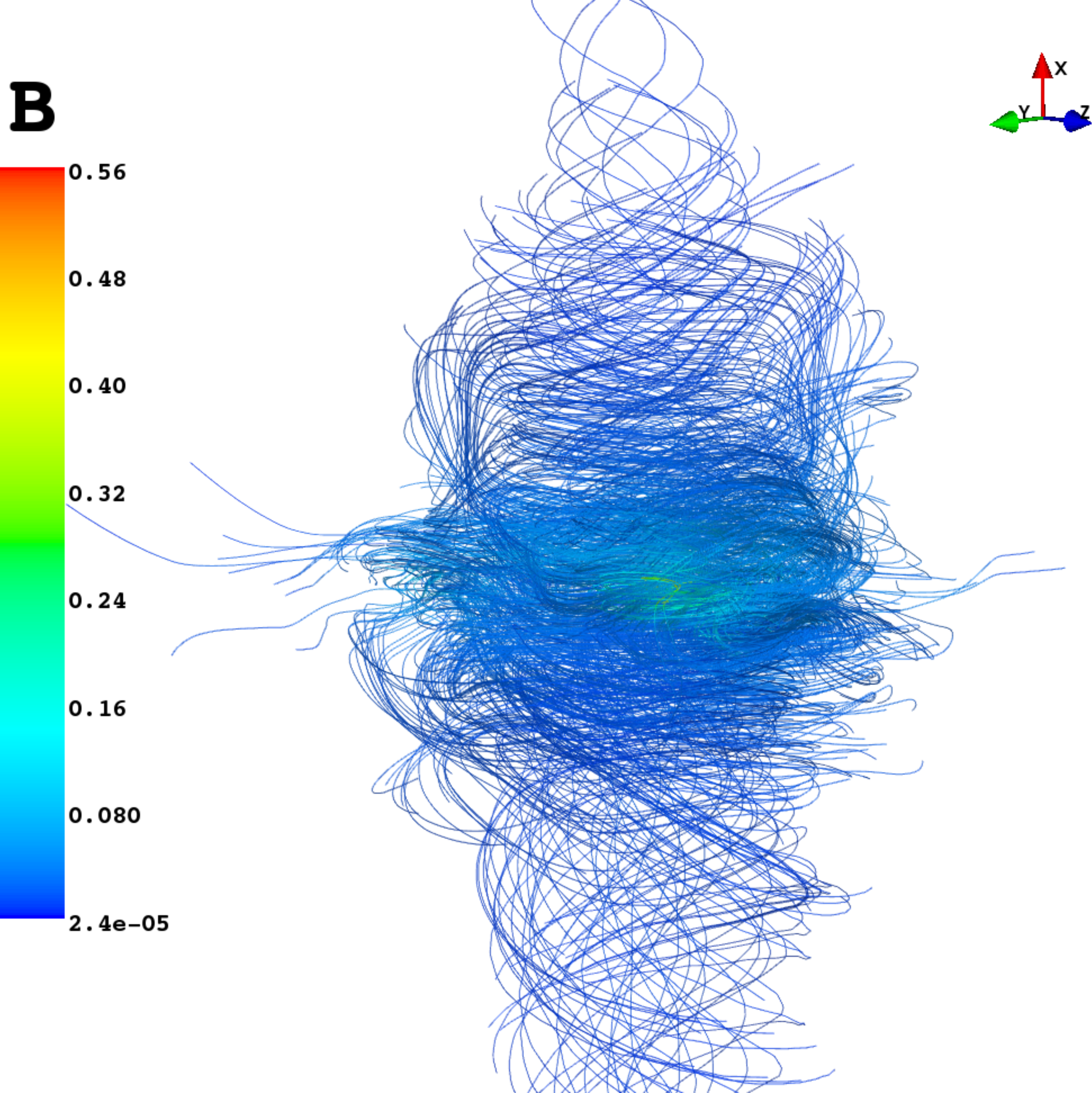}}
\caption[]{CR ionisation maps and iso-density contours ({\em black solid
lines}) for the case E in Table~\ref{lambdaalfatime}. See
Fig.~\ref{RAMSES_M_0.3_0_100} for further information.}
\label{RAMSES_M_0.1_0_44}
\end{figure}

We model the case of weak magnetisation for an aligned rotator and
we consider a late-time configuration (case E in
Table~\ref{lambdaalfatime}).  As shown in the lower panel of
Fig.\ref{RAMSES_M_0.1_0_44}, since the magnetic field braking is
very weak, rotation is able to strongly twist the field lines.  It
is interesting to notice how the contribution to the decrease of
$\zeta^{\rm H_{2}}$ is substantial perpendicularly to the disc
plane.  In fact, the upper right panel of Fig.~\ref{RAMSES_M_0.1_0_44}
shows a large region along the $x$ axis where $\zeta^{\rm
H_{2}}\lesssim10^{-18}$~s$^{-1}$.  This region is not limited to
the high-density domain ($n>10^{9}$~cm$^{-3}$), but it broadens out
along the rotation axis where the magnetic mirroring due to field
line tangling up is very marked. As for case C
(Sect.~\ref{intermediate}), the region with $n\gtrsim10^{9}$~cm$^{-3}$
and $\zeta^{\rm H_{2}}\lesssim10^{-18}$~s$^{-1}$ is characterised by 
$B_{\varphi}/|{\bf B}|\gtrsim0.4$. In the $(y,z)$ plane one can see the presence of
the face-on disc and the rapid decrease of CR ionisation rate that
reaches about $2\times10^{-20}$~s$^{-1}$ in the inner $10^{10}$~cm$^{-3}$
iso-density contour.
This is compatible with the results of Mellon \& Li~(\cite{ml09}) 
who find that a centrifugally supported disc of radius $\sim 50$~AU is formed 
by the collapse of a cloud with $\lambda=13.3$, close to our
value of 17, if $\zeta^{\rm H_{2}}=1\times 10^{-18}$~s$^{-1}$,
which is exactly the value that we find in this region.
Again, the agreement found in this case may not be particularly significant, because in 
clouds characterised by an initially weak field, the magnetic braking 
may result inefficient regardless of the degree of ionisation of the gas. We stress
again that, in order to draw firm conclusions on the role of CR ionisation in the 
resolution of the so-called ``magnetic braking problem'', a fully self-consistent 
calculation including CR propagation and magnetic diffusion is required. 
In Sect.~\ref{fittingformula} we outline a procedure to include approximate 
treatment of CR transport in a MHD simulation.

\section{Cosmic-ray ionisation rate at high column densities}
\label{comparisonUN}

The regime of high column densities, $N({\rm H}_2) > 10^{25}$~cm$^{-2}$,
corresponding to surface mass densities $\Sigma$ larger than a few
g~cm$^{-2}$, is interesting for applications to protoplanetary
discs, where the ionisation rate plays a major role in determining
the ``dead zones'' where the generation of turbulence by the
magnetorotational instability is relatively inefficient (see e.g.
Sano et al.~\cite{sm00}, Okuzumi \& Hirose~\cite{oh11}). In these
applications, the CR ionisation rate is usually assumed to be
exponentially attenuated  within the disc, with a decay length
$\Sigma_0 \approx 96$~g~cm$^{-2}$ (Umebayashi \&
Nakano~\cite{un81})\footnote{The calculations of PGG09 and PGG13
suggest however that the attenuation length for protons may be a
factor of $\sim 2$ larger than that derived by Umebayashi \&
Nakano~(\cite{un81}).}.  Umebayashi \& Nakano~(\cite{un09}) propose
an empirical formula for the CR ionisation rate as a function of
the depth from the disc surface, assuming a geometrically thin disc
and taking into account the fact that CRs penetrate the disc almost
isotropically.  Their formula in cylindrical coordinates reads
\begin{eqnarray}
\label{un}
\zeta^{\rm H_{2}}(r,z)&\approx&\frac{\zeta_{0}^{\rm H_{2}}}{2}%
\Bigg\{\exp\left(-\frac{\Sigma^{+}(r,z)}{\Sigma_{0}}\right)%
\left[1+\left(\frac{\Sigma^{+}(r,z)}{\Sigma_{0}}\right)^{3/4}\right]^{-4/3}\\\nonumber
&&+\ \exp\left(-\frac{\Sigma^{-}(r,z)}{\Sigma_{0}}\right)%
\left[1+\left(\frac{\Sigma^{-}(r,z)}{\Sigma_{0}}\right)^{3/4}\right]^{-4/3}%
\Bigg\}\,,
\end{eqnarray}
where $\zeta_{0}^{\rm H_{2}}\approx10^{-17}$~s$^{-1}$ and
$\Sigma^{+}$ and $\Sigma^{-}$ are the vertical gas surface 
densities measured from the upper ($z_{\rm up}$) 
and the lower boundary ($z_{\rm low}$) of the 
computational domain, respectively, namely
\be
\Sigma^{+}(r,z)=\mu m_{\rm p}\int_{z}^{z_{\rm up}}n(r,z^{\prime})\ \ud z^{\prime}\,,
\ee
and
\be
\Sigma^{-}(r,z)=\mu m_{\rm p}\int_{z_{\rm low}}^{z}n(r,z^{\prime})\ \ud z^{\prime}\,,
\ee
where $m_{\rm p}$ is the proton mass and $\mu=2.36$ is the molecular
weight.  We consider the numerical models described in Sect.~\ref{nummod}
that allow the formation of a keplerian disc (case C and E in
Table~\ref{lambdaalfatime}) comparing our CR ionisation rate maps
with those obtained by using Eq.~(\ref{un}).

\begin{figure*}[t]
\centering
\includegraphics[width=17cm]{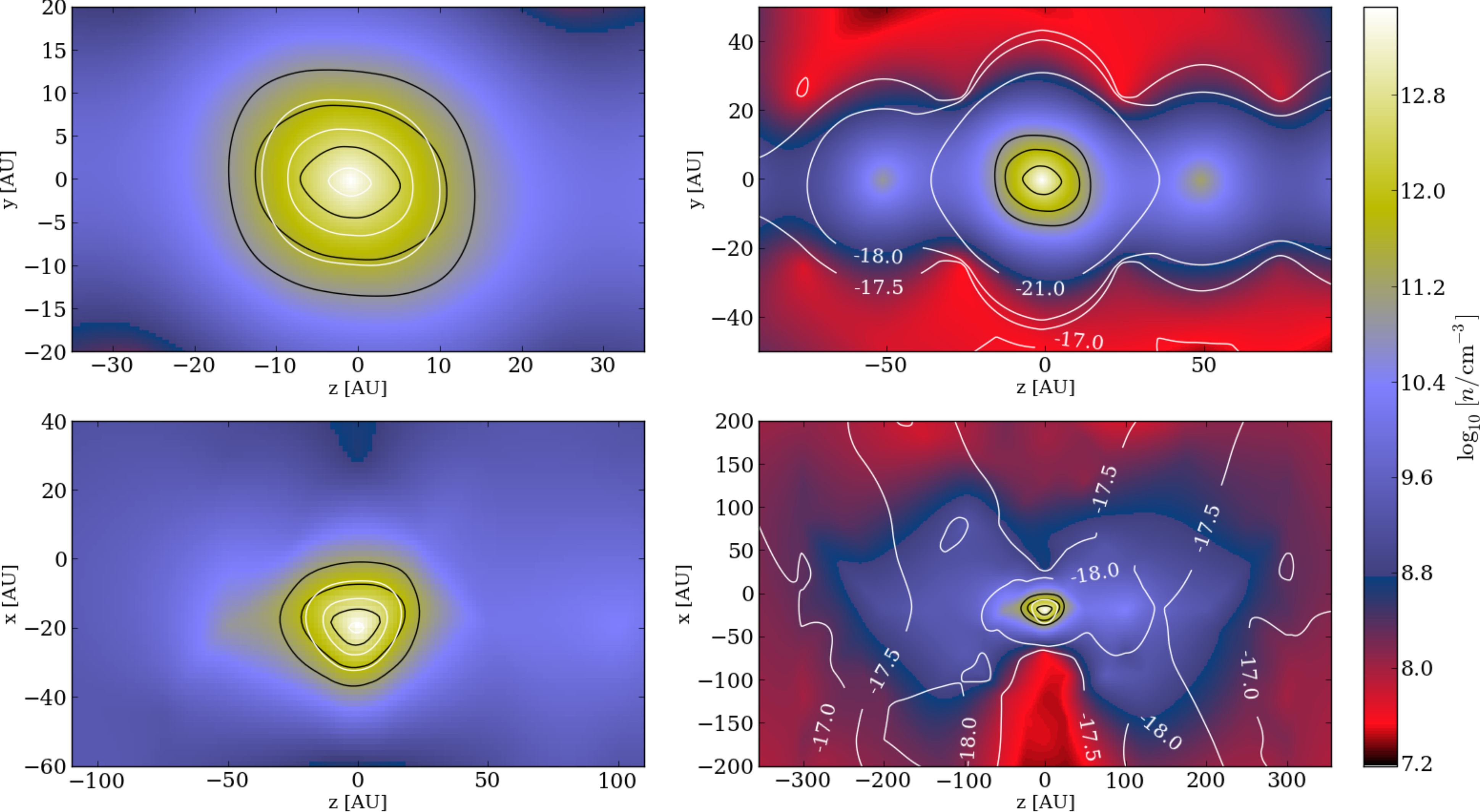}
\caption[]{Comparison between the CR ionisation rate distribution
obtained by using the fitting formula (Eq.~\ref{un}) from Umebayashi
\& Nakano~(\cite{un09}) considering rectilinear propagation, {\em
black contours}, and Eq.~(\ref{fitfor}), {\em white contours}, for
rectilinear propagation ({\em left panels}) and following the path
of CRs along field lines with the inclusion of mirroring and focusing
effects ({\em right panels}).  The three concentric {\em black
contours} in all the panels as well as the {\em white contours} in
the {\em left} panels refer to $\log_{10}\ (\zeta^{\rm H_{2}}/{\rm
s^{-1}})= -17.5, -18, -21$ going inwards. Labels are on a logarithmic
scale and the colour coding shows the density distribution.  {\em
Upper} panels\/: model C; {\em lower} panels\/: model E of
Table~\ref{lambdaalfatime}.}
\label{confrontoUN09}
\end{figure*}

In order to make a consistent comparison, we let CRs propagate along
straight lines without including magnetic effects, computing
$\zeta^{\rm H_{2}}$ using Eq.~(\ref{fitfor}).  This equation gives
similar results to Eq.~(\ref{un}), the small variations being due
to the difference in the assumed attenuation length (see left panels
of Fig.~\ref{confrontoUN09}).  Nevertheless, the picture changes
dramatically when considering the true path of CRs along the field
lines with the inclusion of magnetic and focusing effects. In fact,
the region with $\zeta^{\rm H_{2}}<10^{-18}$~s$^{-1}$ grows
considerably (see right panels
of Fig.~\ref{confrontoUN09}).

This comparison represents the ultimate proof of the role of magnetic
fields on CR propagation. Magnetic effects cannot be neglected since
they set the extent to which CRs can determine the coupling between
magnetic fields and the gas. In the next Section, we give a new
fitting formula for $\zeta^{\rm H_{2}}$ that accounts for the
magnetic configuration.

\section{Magnetic effects on the reduction of the CR ionisation rate: a useful fitting formula}
\label{fittingformula}

In order to compare the relative importance of density and magnetic
effects on the decrease of $\zeta^{\mathrm{H_{2}}}$ in the central
region of a core ($n\gtrsim10^{9}$~cm$^{-3}$), we let CRs propagate
along field lines without accounting for mirroring and focusing,
we refer to this setting as the ``non-magnetic case''.
This is easily done by following the path of CRs with initial pitch
angle $\alpha_{\rm ICM}=0$.  According to Eq.~(\ref{alpha}), the
pitch angle $\alpha$ during the propagation of the particle remains
equal to zero.  We generate CR ionisation maps as those presented
in Figures~\ref{RAMSES_M_0.3_0_100}\ -- \ref{RAMSES_M_0.1_0_44} and
then we compute the quantity ${\cal R}$ defined as the ratio between
the rates obtained including and neglecting magnetic effects as in
PG11.

\begin{figure}[bh!]
\begin{center}
\includegraphics[width=.5\textwidth, clip=true]{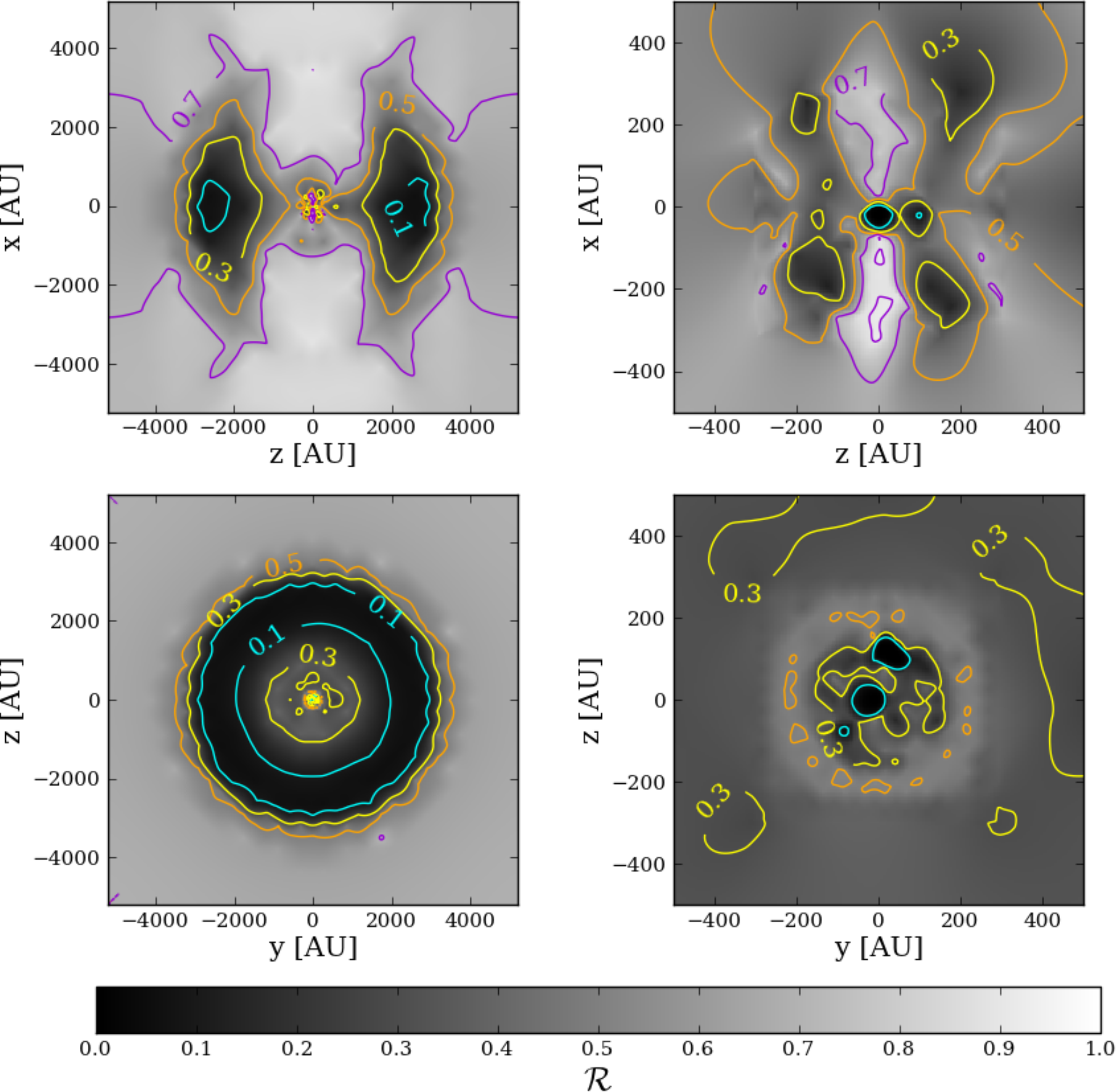}
\caption[]{Maps of the ratio ${\cal R}$ between the CR ionisation rates in the magnetic and non-magnetic 
case for the case E in Table~\ref{lambdaalfatime}. {\em Left} panels show the entire computational domain
while {\em right} panels show a zoom in the inner region.
{\em Upper} panels show a slice of a plane parallel to the 
magnetic field direction, while 
{\em lower} panels refer to a slice of a perpendicular plane. Iso-contours of ${\cal R}$ are shown in 
{\em cyan} (${\cal R}=0.1$), {\em yellow} (${\cal R}=0.3$), {\em orange} (${\cal R}=0.5$), and
{\em purple} (${\cal R}=0.7$).}
\label{rapporto_casoE}
\end{center}
\end{figure}

We focus on cases C and E (see Table~\ref{lambdaalfatime}),
corresponding to the cases where a keplerian disc is formed. In
these models, the density reaches values larger than $\sim
10^{10}$~cm$^{-3}$ inside a region of a few 100 AU in size. Thus,
one can be led to conclude that magnetic effects may be negligible
since CRs have crossed a large column density so that the regime
of exponential attenuation ($N\gtrsim10^{25}$~cm$^{-2}$) is attained.  On
the contrary, we find that even at very high densities the role of
magnetic fields in removing CRs and attenuating $\zeta^{\mathrm{H_{2}}}$
is substantial. The results
are shown in Fig.~\ref{rapporto_casoE}, where we plot the ratio
${\cal R}$ of the CR ionisation rate computed keeping into account
the evolution of the pitch angle to the same quantity computed in
the non-magnetic case.  We see that magnetic effects cause a
net reduction of $\zeta^{\mathrm{H_{2}}}$ of a factor of $\sim 2$
at large scales up to a factor of 3 to 10 in the inner regions.

We give a convenient fitting formula in order to reproduce the
ionisation rate maps without running the full code described in the
paper.  When in presence of a magnetic field, the effective column
density, $N_{\rm eff}$, seen by a CR can be much larger than that
obtained through a rectilinear propagation (see also
Sect.~\ref{comparisonUN}): the more complex the morphology of the
magnetic field, the larger is the path covered by a charged particle.
If $N(\mathrm{H_{2}})$ is the average column density seen by an
isotropic flux of CRs, $N_{\rm eff}$ has the form
\be
\label{Neff}
N_{\rm eff} = (1 + 2\pi\ {\cal F}^{s})\ N(\mathrm{H_{2}})\,.
\ee
The factor ${\cal F}$ depends on the ratio between the toroidal and the poloidal components of
the magnetic field, $b=|B_{\varphi}/B_{p}|$, as well as on its module. It reads
\be\label{calP}
{\cal F}=\frac{|{\bf B}|}{10^{-2}\ \mu{\rm G}}\frac{\sqrt{b^{*}}}{2}\,,
\ee
where
\be\label{betatilde}
b^{*}=\frac{b-b_{\rm min}}{b_{\rm max}-b_{\rm min}}\,,
\ee
$b_{\rm min}$ and $b_{\rm max}$ being the minimum and the maximum
value of $b$ in the whole data cube, respectively.  When the magnetic
field strength is negligible, CRs propagate along straight lines.
In this case ${\cal F}=0$ and $N_{\rm eff}=N({\rm H_{2}})$, otherwise
${\cal F}>0$ and $N_{\rm eff}>N({\rm H_{2}})$.  Comparing the column
density seen by a CR evaluated with our code and with Eq.~(\ref{Neff}),
we find that it is safe to approximate the poloidal configuration
to a rectilinear path. This explains why we introduce the dependence
of ${\cal F}$ on $b^{*}$. In fact, $b^{*}$ varies between 0 and 1,
being equal to 0 for a purely poloidal configuration.  We also find
that the higher the density, the stronger is the role of the magnetic
field in increasing $N_{\rm eff}$. This justifies the presence in
Eq.~(\ref{Neff}) of the power $s$ that reads
\be
\label{spower}
s \simeq 0.7\frac{\log_{10}(n/n_{\rm min})}{\log_{10}(n_{\rm max}/n_{\rm min})}\,,
\ee
$n_{\rm min}$ and $n_{\rm max}$ being the minimum and the maximum
value of the density in the whole data cube, respectively.  The
factor 0.7 in Eq.~(\ref{spower}) has been introduced to reproduce
the CR ionisation maps in the upper and middle panels of
Figures~\ref{RAMSES_M_0.3_0_100}--\ref{RAMSES_M_0.1_0_44}.  Notice
also that the factor ${\cal F}$ depends on the local value of the
magnetic field, since $b$ and $|{\bf B}|$ are computed in a given
point, but it also accounts for the large scale configuration by
means of $b^{*}$ and $s$.

\begin{figure}[!htp]
\begin{center}
\includegraphics[width=.4\textwidth, clip=true]{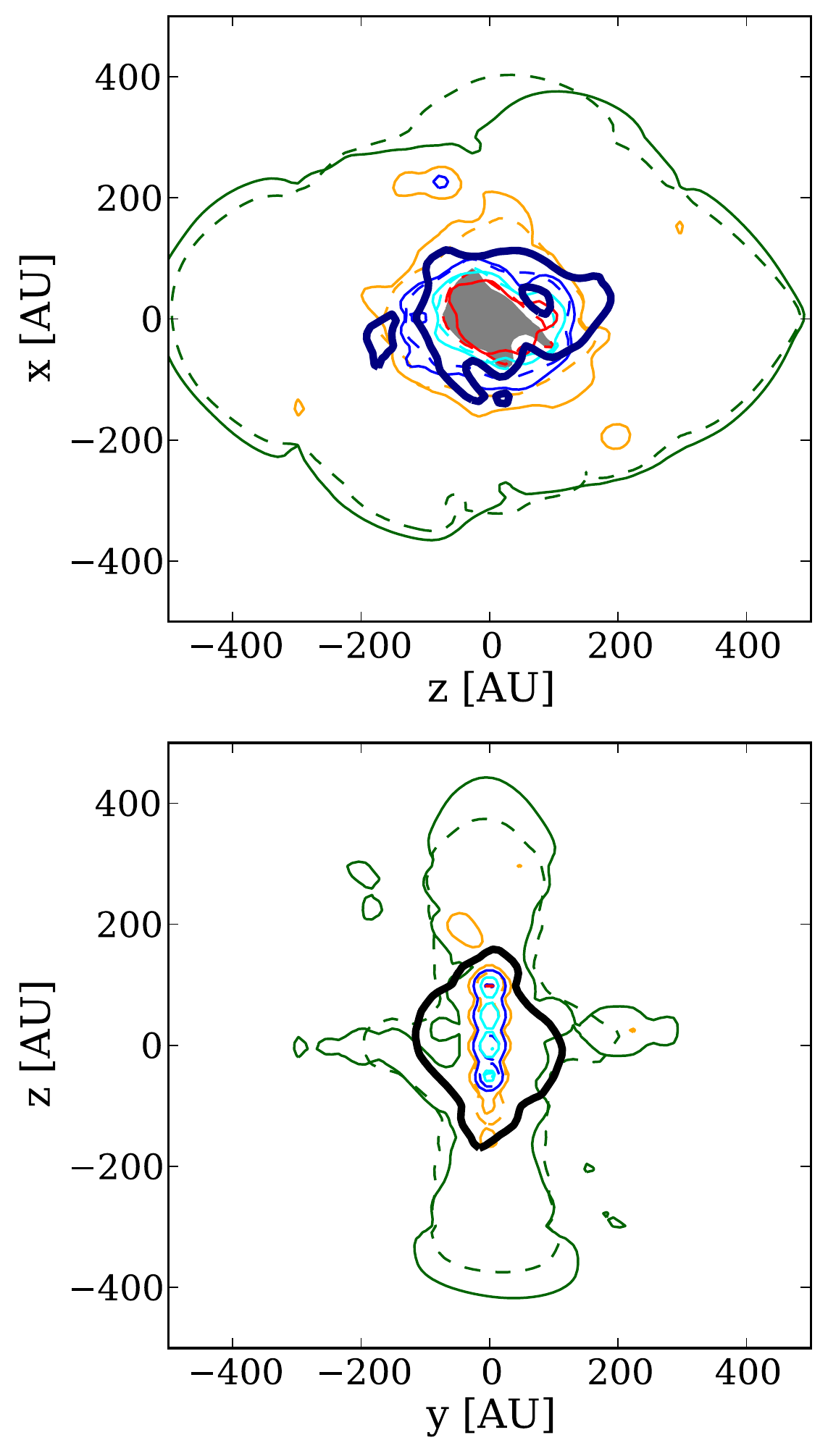}
\caption[]{Comparison between the CR ionisation rate evaluated with the code
described in the paper, {\em solid contours}, and with Eq.~(\ref{zetaeff}),
{\em dashed contours} for the case C in Table~\ref{lambdaalfatime}. 
The {\em grey shaded area} shows the region where the
difference among contours is larger than a factor of 3 and the 
{\em black solid contour} refers to the region where $|{\bf B}|=1$~G.
{\em Green, orange, blue, cyan}, and {\em red contours} are related to
$\log_{10}(\zeta^{\rm H_{2}}/{\rm s}^{-1})=-17, -17.5, -18, -19,$ and $-20$,
respectively.}
\label{fitting_formula_casoC_ver}
\end{center}
\end{figure}

Once evaluated the effective column density, the corresponding
(effective) CR ionisation rate, $\zeta^{\rm H_{2}}_{\rm eff}$, is
obtained by
\be
\label{zetaeff}
\zeta^{\rm H_{2}}_{\rm eff}=\kappa \zeta^{\rm H_{2}}(N_{\rm eff})\,,
\ee
where $\zeta^{\rm H_{2}}(N_{\rm eff})$ is computed using
Eq.~(\ref{fitfor}) after replacing $N(\alpha)$ and $\Sigma(\alpha)$
with $N_{\rm eff}$ and $\mu m_{\rm H}N_{\rm eff}$, respectively.
The factor $\kappa$ is given by
\be\label{kappa}
\kappa = \frac{1}{2}+\frac{1}{\pi}\arctan\left(\frac{10\ \mu{\rm G}}{|{\bf B}|}\right)
\ee
and it represents the correction for magnetic effects.
Equation~(\ref{zetaeff}) gives a correct result within a factor of
3 and it holds for magnetic field strengths smaller than about 1~G.
It can be very helpful for non-ideal MHD simulations so as to compute
diffusion coefficients and it can be also implemented in chemical models
in order to have a more precise description of the observational results.
Figure~\ref{fitting_formula_casoC_ver}
shows the goodness of the fit for the inner region of case C in
Table~\ref{lambdaalfatime}.

\section{Conclusions}
\label{conclusions}

In this study we explored the distribution of the CR ionisation
rate in molecular clouds, examining and extending the results
obtained in PGG09 and PG11. We employed a semi-analytical model in
order to understand when the variations of $\zeta^{\rm H_{2}}$ can
be attributed to energy losses due to the increasing column density
passed through or to magnetic mirroring and focusing.  The main
conclusion is that an increment of the toroidal component, and in
general a more tangled magnetic field, corresponds to a decrease
of $\zeta^{\rm H_{2}}$ because of the growing preponderance of
mirroring over focusing. That is to say, large variations of the
direction of field lines cause a rapid increase of the cosmic-ray
pitch angle towards the mirroring angle.  Conversely, fixing a
magnetic field line configuration while varying the density profile
allowed us to identify the weakening of the ionisation power
caused by energy losses. As expected, moving from a ``roundish''
core towards a disc-like configuration, the distribution of $\zeta^{\rm
H_{2}}$ follows the density profile shape.

In the second part of the paper we analysed a number of numerical
simulation outputs related to a rotating collapsing core following
the propagation of cosmic rays at different time steps, varying the
degree of magnetisation and the initial orientation of the main
magnetic field direction with respect to the rotation axis.
Being aware of the fact
that the correct manner of dealing with CR propagation should be
computing their distribution simultaneously with the MHD simulation,
we believe that our conclusions represent an important proof of
concept.  In particular, in the central 100~AU region, the number
density is higher than $10^{10}$~cm$^{-3}$ and a H$_{2}$ column
density larger than 10$^{25}$~cm$^{-2}$ is promptly reached. This
actually means entering the exponential attenuation regime where
the CR ionisation rate is independent of the CR spectrum assumed
and drops below $10^{-18}$~s$^{-1}$. However, we prove that even
in the high-density region, the presence of a magnetic field can
reduce $\zeta^{\rm H_{2}}$ up to a factor larger than 10.  As for the
semi-analytical model, we also conclude that the morphology of the
$\zeta^{\rm H_{2}}$ maps depends both on the density profile and
on the magnetic field line configuration.  

Finally we focused on the morphology of $\zeta^{\rm H_{2}}$ in the
inner region where a keplerian disc is formed. We found that the
inclusion of magnetic field effects in the calculation of $\zeta^{\rm
H_{2}}$ brings to the formation of a large central region of 100
to 200~AU where the CR ionisation rate is well below the ordinarily
used value of $10^{-17}$~s$^{-1}$.  This provides support to the
hypothesis of Mellon \& Li~(\cite{ml09}) according to which the
magnetic braking efficiency can be reduced if $\zeta^{\rm
H_{2}}\lesssim10^{-18}$~s$^{-1}$ allowing the formation of the disc.
In order to test this hypothesis a self-consistent MHD collapse
calculation including CR propagation is needed.

We compared our results with those obtained by Umebayashi \&
Nakano~(\cite{un09}) in the case of an unmagnetised disc,
showing that the exclusion of CRs 
resulting from magnetic mirroring deeply affects the CR ionisation rate
pattern in the collapse region. To account for the effects of the 
magnetic configuration, we formulated a general fitting expression to 
approximately compute $\zeta^{\rm H_{2}}$ as a function of the 
column density, magnetic field strength, and toroidal-to-poloidal magnetic field
ratio. This empirical expression reproduces quite accurately our numerical
results.

Non-ideal MHD models predict that 
magnetic braking becomes inefficient at densities $n>10^{12}$~cm$^{-3}$, 
when magnetic field diffusion becomes faster than the dynamical evolution
(Dapp et al.~\cite{db12}). In our models we observe that the drop in 
$\zeta^{\rm H_{2}}$ takes place in some cases
even at lower densities ($n>10^{9}$~cm$^{-3}$), resulting in 
very low ionisation fractions. The consequences of the reduced CR ionisation
rate on the magnetic diffusion coefficients (ambipolar, Hall, and Ohm)
will be the subject in a forthcoming paper.  


\acknowledgements
MP thanks Marc Joos and Andrea Ciardi for their help in accessing
RAMSES simulations and Natalia Dzyurkevich for illuminating discussions
about ionisation in discs.  MP and PH acknowledge the financial
support of the Agence National pour la Recherche (ANR) through the
COSMIS project.


\appendix

\section{Models $\cL$ and $\cH$}\label{caseLandH}

\begin{figure}[t]
\centering
\includegraphics[width=.5\textwidth, clip=true]{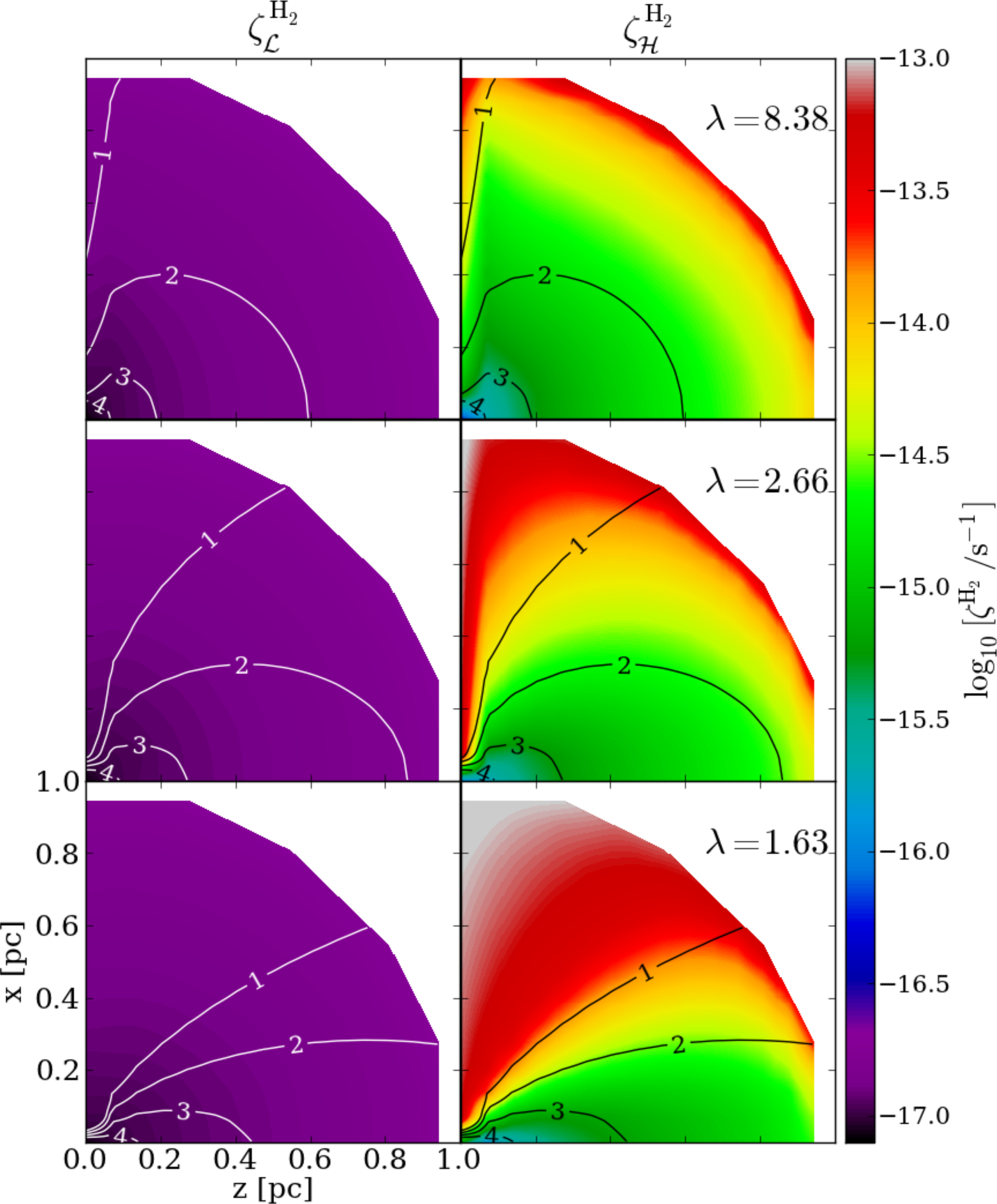}
\caption[]{CR ionisation rate maps for the model $\cL$ ({\em left}
column) and $\cH$ ({\em right} column) in the plane $y=0$ for a
fixed toroidal-to-poloidal ratio $b_{0}=10$ and different values
of the mass-to-flux-ratio.  {\em White and black contours} represent
the iso-density contours and the labels show $\log_{10}\
[n/\mathrm{cm^{-3}}]$.} \label{PLOTanalyticRESULTS_3x3_appendix}
\end{figure}

\begin{figure*}[]
\centering
\includegraphics[width=17cm, clip=true]{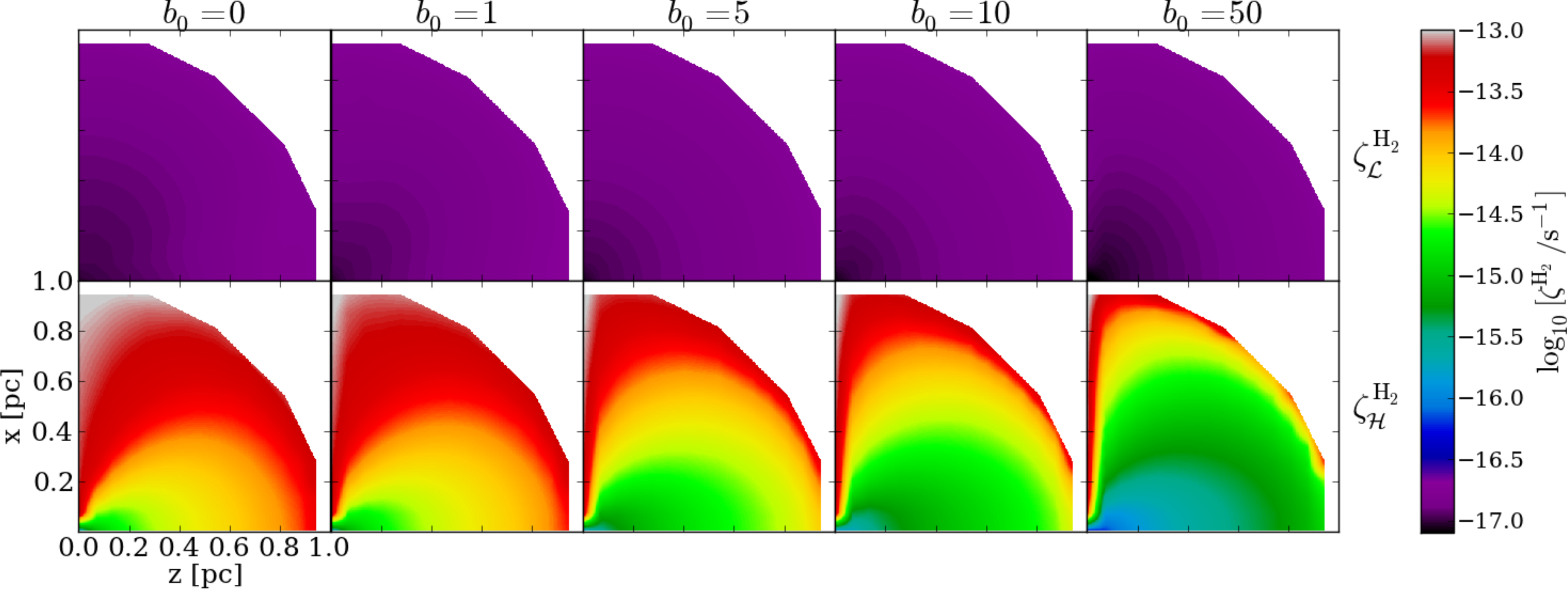}
\caption[]{CR ionisation rate profiles for the model $\cL$ ({\em
top} row) and $\cH$ ({\em bottom} row) in the plane crossing the
symmetry axis and perpendicular to the midplane $(y=0)$. The
mass-to-flux ratio is $\lambda=2.66$ and the strength of the toroidal
field $b_{0}$ increases from left to right.}
\label{PLOTanalyticRESULTS_appendix}
\end{figure*}

For the sake of completeness, we show the CR ionisation rate
$\zeta^{\rm H_{2}}_{k}$ obtained adopting the spectra corresponding
to the cases $k=\cL$ and $k=\cH$ in Fig.~\ref{modelfitsAMRpaper}.
These represent two extreme behaviours of CR ionisation as a function
of column density.  For both cases, we compute the dependence of
$\zeta^{\rm H_{2}}_{k}$ on the density profile and the magnetic
field configuration for the semi-analytical cloud models described
in Sect.~\ref{semi-analytical}.  In particular we show results for
$\lambda=8.38, 2.66$, and 1.63 with $b_{0}=10$
(Fig.~\ref{PLOTanalyticRESULTS_3x3_appendix}) and for $\lambda=2.66$
with $b_{0}=0,1,5,10$, and 50 (Fig.~\ref{PLOTanalyticRESULTS_appendix}).
As expected, the CR ionisation rate for the case $k=\cL$ is independent
of the value of the mass-to-flux ratio and the toroidal-to-poloidal
ratio since the exponential regime ($N({\rm
H_{2}})\gtrsim10^{25}$~cm$^{-2}$) is not attained and $\zeta^{\rm
H_{2}}_{\cL}$ is substantially independent of the column density.

On the other hand, the CR ionisation rate for the model $\cH$ is
comparable with that of the model $\cM$ described in the main text.
However, as shown in Fig.~\ref{PLOTanalyticRESULTS_appendix}, even
when the toroidal field is strong ($b_{0}=50$), in the inner region
of the toroid below 0.1~pc, $\zeta^{\rm H_{2}}_{\cH}$ is of the
order of $10^{-16}$~s$^{-1}$, about one order of magnitude higher
than the values estimated from observations. For this reason, the
model $\cM$ has been chosen as the fiducial model.

\end{document}